\documentclass[11pt,english,aps,pra,onecolumn,tightenlines,groupedaddress,showkeys,notitlepage,floatfix]{revtex4-1}

\usepackage[T1]{fontenc}
\usepackage{amsmath}
\usepackage{amssymb}
\usepackage{amsfonts}
\usepackage{bm}
\usepackage{bbm}
\usepackage{epsfig}
\usepackage{grffile}
\usepackage{times}
\usepackage{amsthm}

\usepackage[usenames,dvipsnames]{color}
\definecolor{dblue}{rgb}{0,0.1,.6}

\usepackage[colorlinks=true,citecolor=dblue,linkcolor=dblue,urlcolor=dblue]{hyperref}
\usepackage[all]{hypcap}

\newcommand{\id}{\mathbbm{1}}
\newcommand{\floor}{\operatorname{floor}}
\newcommand{\ceil}{\operatorname{ceil}}

\newcommand{\mc}[1]{\mathcal{#1}}
\newcommand{\ts}{\textstyle}

\newcommand{\tN}{\text{N}}
\newcommand{\tS}{\text{S}}
\newcommand{\tL}{\text{L}}
\newcommand{\tSL}{\text{SL}}
\newcommand{\tH}{\text{H}}
\newcommand{\tY}{\text{Y}}
\newcommand{\tZ}{\text{Z}}
\newcommand{\tE}{{\text{E}}}
\newcommand{\tEp}{{\text{E}+}}
\newcommand{\tEm}{{\text{E}-}}
\newcommand{\tSE}{\text{SE}}
\newcommand{\veps}{\varepsilon}
\renewcommand{\vec}[1]{{\boldsymbol{#1}}}

\newcommand{\myitem}{{\textbullet \ }}
\newcommand{\myitemBest}{{\scalebox{0.8}{$\bigstar$}\ }}
\newcommand{\myitemSummary}{{\scalebox{0.7}{$\blacksquare$} \ }}
\newcommand{\imagetop}[1]{\vtop{\null\hbox{#1}}}

\newtheoremstyle{thmstyle}
  {0.1em} 
  {0em} 
  {} 
  {} 
  {\bfseries} 
  {.} 
  {.5em} 
  {} 
\theoremstyle{thmstyle} 
\newtheorem{lemma}{Lemma}
\newcommand{\lemmaHead}[1]{\bfseries \emph{#1}}

\makeatletter
\renewcommand{\p@subsection}{}
\renewcommand{\p@subsubsection}{}
\makeatother

\newcommand{\duke} {Department of Physics, Duke University, Durham, North Carolina 27708, USA}

\newcommand{\Title} {Optimized Lie-Trotter-Suzuki decompositions for two and three non-commuting terms}
\newcommand{\Authors}
{
\author{Thomas Barthel and Yikang Zhang\vspace{-0.5em}}
\affiliation{\duke}
}
\newcommand{\Date} {March 03, 2019}

\begin{document}

\title{\Title}
\Authors

\begin{abstract}
Lie-Trotter-Suzuki decompositions are an efficient way to approximate operator exponentials $\exp(t H)$ when $H$ is a sum of $n$ (non-commuting) terms which, individually, can be exponentiated easily. They are employed in time-evolution algorithms for tensor network states, digital quantum simulation protocols, path integral methods like quantum Monte Carlo, and splitting methods for symplectic integrators in classical Hamiltonian systems. 
We provide optimized decompositions up to order $t^6$. The leading error term is expanded in nested commutators (Hall bases) and we minimize the 1-norm of the coefficients. For $n=2$ terms, several of the optima we find are close to those in McLachlan, SlAM J.\ Sci.\ Comput.\ \textbf{16}, 151 (1995). Generally, our results substantially improve over unoptimized decompositions by Forest, Ruth, Yoshida, and Suzuki. We explain why these decompositions are sufficient to efficiently simulate any one- or two-dimensional lattice model with finite-range interactions. This follows by solving a partitioning problem for the interaction graph.
\end{abstract}

\date{\Date\vspace{-0.5em}}

\keywords{Lie-Trotter-Suzuki decomposition, Lie-Trotter product formula, operator exponential, tensor network states, quantum Hamiltonian simulation, quantum Monte Carlo, symplectic integrators, splitting methods, Baker-Campbell-Hausdorff formula, Hall basis, Gr\"{o}bner basis, graph partitioning problem}

\maketitle
\vspace{-2.5em}
\renewcommand{\baselinestretch}{0.85}\normalsize
\tableofcontents
\renewcommand{\baselinestretch}{1}\normalsize

\section{Introduction}\label{sec:intro}
In many situations, we need to evaluate or apply operator exponentials $e^{t H}$ where $H$ acts in a huge vector space. A first trick is to decompose the task into small time steps, $e^{(t_2-t_1) H}=(e^{t H})^N$ with $N t = t_2-t_1$, and we will always consider $t$ to denote that time step in the following. While $H$ refers to a Hamiltonian in many applications, we do not assume $H$ to be Hermitian or skew-Hermitian. For example, it could also be a Liouville super-operator for a Markovian open quantum system.
When $H$ is a sum of $n$ (non-commuting) terms which, individually, can be exponentiated easily, one can approximate $e^{t H}$ by a product of these easy exponentials \cite{Trotter1959,Suzuki1976-51}. This gives Lie-Trotter-Suzuki decompositions like
\begin{equation}\label{eq:LTSintro}
	e^{t H}=e^{x_1 tA}\,e^{x_2 tB}\,e^{x_3 tA}\,e^{x_4 tB}\dotsb e^{x_m tA} + \mc{O}(t^{p+1})
\end{equation}
for the case of a generator $H=A+B$ with $n=2$ terms. Depending on the number of factors $m$ in the product approximation \eqref{eq:LTSintro}, one can achieve different approximation orders $p$ with respect to $t$ and may still have free parameters to minimize the approximation error. In particular, when there are free parameters left, we can use them to minimize the amplitude of the leading order error term $\propto t^{p+1}$.
We discuss and present such optimized Lie-Trotter-Suzuki decompositions for $n=2$ and $n=3$ terms in the generator $H$.

Lie-Trotter-Suzuki decomposition have many applications. For example, they are employed in time evolution algorithms for matrix product states \cite{Vidal2003-10,White2004,Daley2004,Orus2008-78} or higher-dimensional tensor network states called projected entangled pair states \cite{Verstraete2004-7,Verstraete2006-96,Niggemann1997-104,Nishino2000-575,Martin-Delgado2001-64}. Importantly, Lie-Trotter-Suzuki decompositions can also be used for digital quantum simulation of many-body systems on quantum computers \cite{Lloyd1996-273,Barry2007-270,Lanyon2011-334,Kliesch2011-107,Barreiro2011-470}.
Furthermore, they are important tools in path integral methods like worldline quantum Monte Carlo \cite{Suzuki1977-58}, diffusion Monte Carlo \cite{Kolorenc2011-74}, and for approximate symplectic integrators for the Hamilton equations of classical systems \cite{Ruth1983-30,Yoshida1990,McLachlan1995}. In the latter context, the decompositions are also called splitting or composition methods.

Here, we provide optimized decompositions up to order $p=6$ for $n=2$ and $n=3$ terms in the generator $H$. We minimize a bound on the operator norm of the leading error term by expanding in terms of nested commutators (Hall bases) and optimizing the 1-norm of the coefficients. In the comparison of different decompositions, we take into account that increased numbers $m$ of factors in the decomposition need to be compensated by correspondingly larger time steps $t$ to keep computation costs constant.
For $n=2$ terms, with a few exceptions, we find optima that are close to those in Ref.~\cite{McLachlan1995}, where McLachlan minimized an error measure suitable for classical Hamiltonian systems. Generally, our results improve substantially over unoptimized decompositions due to Forest and Ruth, Yoshida, and Suzuki \cite{Forest1990-43,Yoshida1990,Suzuki1990}. We also compare to decompositions of Kahan and Li, and Omelyan et al.\ \cite{Kahan1997-66,Omelyan2002-146}. The optimized decompositions for $n\in\{2,3\}$ can be used to efficiently simulate any one-dimensional (1d) and two-dimensional lattice model with finite-range interactions (geometrically local Hamiltonians) as we explain with a coarse-graining argument. In fact, several of the considered decompositions (those of type SL and SE) are generally applicable for any number of terms $n$, but they are in general not optimal when applied for $n\geq 4$.

In section~\ref{sec:coarseGrain}, we show that interaction graphs of 1d and 2d lattice models with finite-range interactions can always be partitioned into $n\leq 3$ subgraphs $A$, $B$ (and $C$), each corresponding to sums of local operators with disjoint spatial supports.
In section~\ref{sec:n2}, we discuss Lie-Trotter-Suzuki decompositions for $n=2$ terms and further general aspects: the considered types of decompositions (Sec.~\ref{sec:n2Decomps}), the Baker-Campbell-Hausdorff formula which is essential to expand decompositions in terms of nested commutators (Sec.~\ref{sec:BCH}), Hall bases for the free Lie algebra generated by the terms in $H$ (used to remove linear dependence of nested commutators; see Sec.~\ref{sec:n2HallBases}), the numbers of parameters and constraints in the different types of $n=2$ decompositions (Sec.~\ref{sec:n2NoParam}), Gr\"{o}bner bases which can be used to solve the systems of polynomial constraints and to identify free parameters (Sec.~\ref{sec:Groebner}), the 1-norm error measure which bounds the operator-norm approximation accuracy and is minimized in optimizations as well as the issue of quasi-locality (Sec.~\ref{sec:n2ErrorMeasure}), and generic unoptimized decompositions due to Forest, Ruth, Yoshida, and Suzuki (Sec.~\ref{sec:n2Unoptimized}).
In section~\ref{sec:n2opt}, we provide various optimized decompositions for $n=2$, compare to previous results, and make recommendations.
Section~\ref{sec:n3} describes the considered types of decompositions for $n=3$ terms (Sec.~\ref{sec:n3Decomps}) as well as their numbers of parameters and constraints (Sec.~\ref{sec:n3NoParam}).
In section~\ref{sec:n3opt}, we provide various optimized decompositions for $n=3$, compare to previous results, and make recommendations.
We close with a short discussion of the results and alternatives to Lie-Trotter-Suzuki decompositions (Sec.~\ref{sec:discuss}).

\section{Coarse-graining and partitioning of interaction graphs}\label{sec:coarseGrain}
\begin{figure}[t]
	\includegraphics[width=0.98\columnwidth]{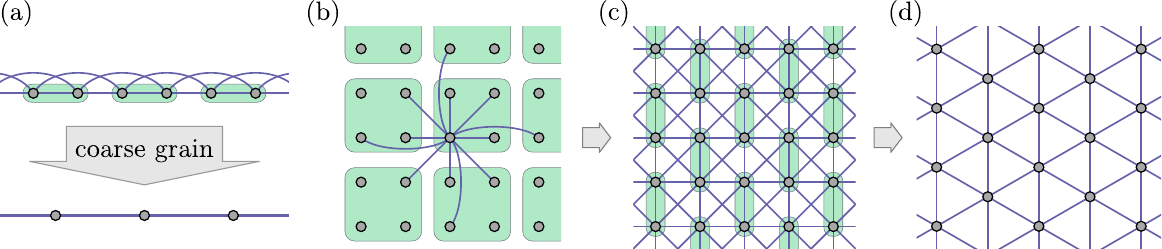}
	\caption{\label{fig:coarseGrain}\textbf{(a)} Coarse graining transforms 1d lattice models with finite-range interactions to models with nearest neighbor interactions only. The diagram shows a system with nearest and next-nearest neighbor interactions (blue lines), and coarse graining of two neighboring sites into one as indicated by green boxes. \textbf{(b-d)} We can use lattice deformations and coarse graining to transform any 2d lattice model with finite-range interactions into a model with nearest neighbor interactions on a triangular lattice.}
\end{figure}
The first step in the design of Lie-Trotter-Suzuki decompositions for $e^{t H}$ is to partition $H$ into a sum of $n$ terms $A,B,\dots$ such that each of these terms can be exponentiated easily, i.e., with low cost. For classical systems, these are typically the kinetic energy term and the potential energy term.
In this section, we consider quantum many-body lattice models with finite-range 2-local interactions. As explained in the following, we can always find partitionings into $n\leq 2$ terms for such 1d systems and $n\leq 3$ terms for 2d systems. In particular, every term will be a sum of mutually commuting operators with finite (local) spatial support. As all elementary interactions are two-particle interactions, 2-locality is natural and, in solid state systems, lattice models with finite-range interactions naturally arise due to screening effects and the tight-binding approximation \cite{Mahan2000}.

Using coarse graining as in Kadanoff's block spin transformation, we can always reduce the problem to nearest neighbor interactions:
\begin{lemma}[\lemmaHead{Reduction to nearest neighbor interactions}]\label{lemma:nearestNeighbor}
  Consider lattice models with finite-range 2-local interactions. In 1d, a finite number of coarse-graining steps is sufficient to map to a model with onsite and nearest neighbor interactions only. In 2d, a finite number of coarse-graining steps is sufficient to map to a model on the triangular lattice with onsite and nearest neighbor interactions only.
\end{lemma}
The situation for 1d is displayed in Fig.~\ref{fig:coarseGrain}(a). When we coarse grain two sites into one new effective site, an interaction between sites of distance $\Delta x$ maps to an interaction between effective sites of distance $\floor(\Delta x/2)$ or $\floor(\Delta x/2)+1$. After a finite number of coarse-graining steps, we arrive at a model with interactions of range $\Delta x\in\{0,1\}$.
The situation for 2d is displayed in Fig.~\ref{fig:coarseGrain}(b-d). In a first step, any 2d lattice can be deformed into a square lattice. When we then coarse grain a square of four sites into one new effective site, an interaction between sites of distance $(\Delta x,\Delta y)$ maps to an interaction between effective sites with $x$ distance $\floor(\Delta x/2)$ or $\floor(\Delta x/2)+1$ and $y$ distance $\floor(\Delta y/2)$ or $\floor(\Delta y/2)+1$. After a finite number of coarse-graining steps, we arrive at a square lattice with interactions of range $\Delta x,\Delta y\in\{0,1\}$, i.e., nearest and next-nearest neighbor interactions as shown in Fig.~\ref{fig:coarseGrain}(c). In a final coarse graining step, we can map two neighboring sites to one effective site to arrive at a triangular lattice with onsite and nearest neighbor interactions only [Fig.~\ref{fig:coarseGrain}(c,d)].

\begin{figure}[t]
	\includegraphics[width=0.98\columnwidth]{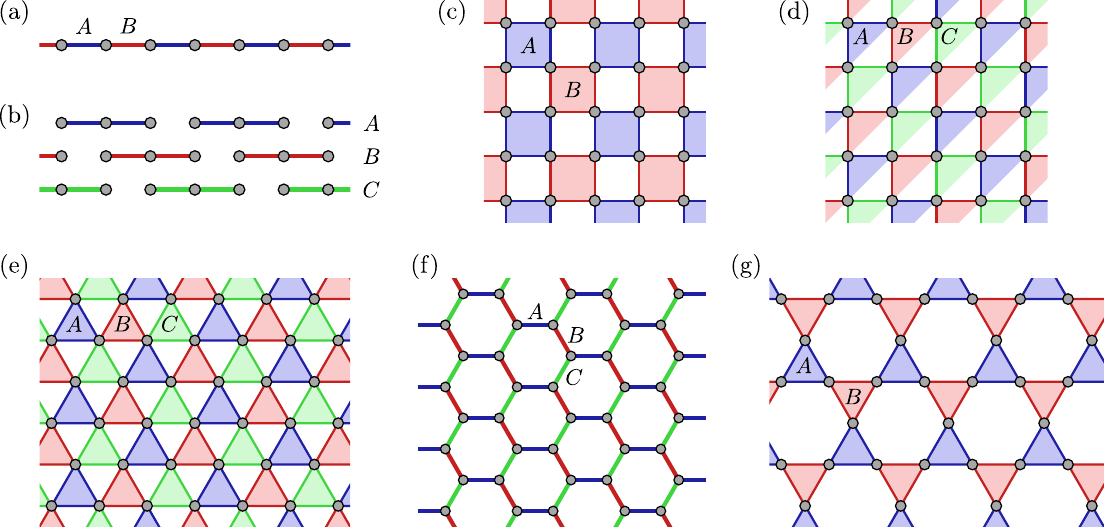}
	\caption{\label{fig:partition}\textbf{(a)} For 1d lattice models with nearest neighbor interactions, the Hamiltonian (viz.\ the interaction graph) can be partitioned into $n=2$ terms $A$ and $B$, each consisting of commuting two-site operators. \textbf{(b)} For 1d lattices with nearest and next-nearest neighbor interactions, one could coarse grain once to then partition into $n=2$ terms, each consisting of four-site operators. Alternatively, we can partition into $n=3$ terms $A$, $B$, and $C$, each consisting of commuting three-site operators. \textbf{(c,d)} For a square lattice, nearest neighbor interactions can be partitioned into two (three) terms, each consisting of commuting four-site (three-site) operators. \textbf{(e,f)} For triangular and hexagonal lattices (tilings) with nearest neighbor interactions, we can partition into three terms with three-site and two-site operators, respectively. \textbf{(g)} For the Kagom\'{e} lattice, two terms  are sufficient, both consisting of commuting three-site operators.}
\end{figure}

\begin{lemma}[\lemmaHead{Partitioning 1d and 2d interaction graphs}]\label{lemma:partition}
  Consider lattice models with finite-range 2-local interactions. The interactions for any 1d model can be partitioned into $n=2$ terms. The interactions for any 2d model can be partitioned into $n=3$ terms. Each of the resulting terms is a sum of local
  interaction operators with disjoint spatial supports.
\end{lemma}
According to Lemma~\ref{lemma:nearestNeighbor}, we can use coarse graining to reduce the problem to nearest neighbor interactions on 1d lattices and 2d triangular lattices. For 1d lattices with nearest neighbor interactions, the interaction graph can be partitioned into two terms, the first containing interactions on even bonds (of effective sites after coarse graining) and the second containing interactions on odd bonds as shown in Fig.~\ref{fig:partition}(a). For 2d triangular lattices with nearest neighbor interactions, the interaction graph can be partitioned into three terms. All interactions are grouped into three-site operators acting on triangular plaquettes. As indicated in Fig.~\ref{fig:partition}(e), three terms ($A,B,C$) are required such that the three-site operators in each term are all mutually commuting.

The described coarse-graining to achieve a partitioning into $n=2$ or $3$ terms, is always possible in 1d and 2d but may not always be the best choice. For instance, in a 1d model with nearest and next-nearest neighbor interactions, we would coarse grain once [Fig.~\ref{fig:coarseGrain}(a)] to arrive at a partitioning with $n=2$. The resulting nearest neighbor interactions between effective sites correspond to four-site operators for the original lattice. Depending on the specific classical or quantum computation costs, it may be preferable to partition into three terms as in Fig.~\ref{fig:partition}(b) and use an $n=3$ Lie-Trotter-Suzuki decomposition instead of a decomposition for $n=2$. Another example is the square 2d lattice with nearest neighbor interactions. One can partition into $n=2$ terms with four-site operators [Fig.~\ref{fig:partition}(c)] or into $n=3$ terms with three-site operators [Fig.~\ref{fig:partition}(d)].

\section{Lie-Trotter-Suzuki decompositions for \texorpdfstring{$n=2$}{n=2} terms and general aspects}\label{sec:n2}
For $n=2$, the generator $H=A+B$ consists of two terms, where exponentials $e^{tA}$ and $e^{tB}$ can be computed easily. For example, both can be sums of commuting few-particle interaction terms. Examples for lattice models are given in Fig.~\ref{fig:partition}(a,c,g). Keep in mind that we actually do not assume $H$ to be Hermitian such that our decompositions apply equally well to real- and imaginary-time evolution, the evolution under Liouvillians, and quite generally, for any matrix exponentials.

\subsection{Considered types of decompositions}\label{sec:n2Decomps}
We consider the following types of Lie-Trotter-Suzuki decompositions for $e^{tH}$, where $m$ denotes the total number of operator exponentials:
\begin{itemize}
  \item
  \textbf{Type N} with $\nu=m$ parameters. This is the most generic type of decompositions
  \begin{equation}\label{eq:n2TypeN}
  	U_{\tN,m}:=
  	\begin{cases}
  	e^{a_1 tA}\,e^{b_1 tB}\,e^{a_2 tA}\,e^{b_2 tB}\dotsb e^{b_{m/2} tB} &\text{for even $m$},\\
  	e^{a_1 tA}\,e^{b_1 tB}\,e^{a_2 tA}\,e^{b_2 tB}\dotsb e^{a_{(m+1)/2} tA} &\text{for odd $m$}
  	\end{cases}
  \end{equation}
  with $a_i,b_i\in\mathbb{R}$.
  \item
  \textbf{Type S} with $\nu=(m+1)/2$ parameters and odd $m$. This is the most generic type of symmetric decompositions
  \begin{equation}\label{eq:n2TypeS}
  	U_{\tS,m}:=
  	\begin{cases}
  	e^{a_1 tA}\,e^{b_1 tB}\,e^{a_2 tA}\dotsb e^{b_{\nu-1} tB}\,e^{a_{\nu} tA}\,e^{b_{\nu-1} tB} \dotsb e^{a_1 tA} &\text{for odd $\nu$},\\
  	e^{a_1 tA}\,e^{b_1 tB}\,e^{a_2 tA}\dotsb e^{a_{\nu-1} tA}\,e^{b_{\nu} tB}\,e^{a_{\nu-1} tA} \dotsb e^{a_1 tA} &\text{for even $\nu$}
  	\end{cases}
  \end{equation}
  with $a_i,b_i\in\mathbb{R}$.
 \item
  \textbf{Type SL} with $\nu=\ceil((m-1)/4)$ parameters and odd $m$. This type of symmetric decomposition is a product of ``leapfrog'' terms \footnote{The name alludes to the similarity to the leapfrog integration method for differential equations. This second order decomposition is also known as the Verlet integrator.}
  \begin{equation}\label{eq:leapfrog}
  	U_{\tL}(\tau):= e^{\frac{1}{2}\tau tA}\,e^{\tau B}\,e^{\frac{1}{2}\tau A}
  \end{equation}
   such that 
  \begin{align}\nonumber
  	U_{\tSL,m}&:=
  	 \begin{cases}
  	  U_\tL(w_1t)U_\tL(w_2t)\dotsb U_\tL(w_{\nu-1}t)U_\tL(w_{\nu}t)U_\tL(w_{\nu-1}t) \dotsb U_\tL(w_1t) &\text{for odd $\frac{m-1}{2}$},\\
  	  U_\tL(w_1t)U_\tL(w_2t)\dotsb U_\tL(w_{\nu-1}t)U_\tL(w_{\nu}t)U_\tL(w_{\nu}t) \dotsb U_\tL(w_1t) &\text{for even $\frac{m-1}{2}$}
  	 \end{cases}\\
  	&\phantom{:}=e^{\frac{1}{2}w_1 tA}\,e^{w_1 tB}\,e^{\frac{1}{2}(w_1+w_2) tA}\,e^{w_2 tB}\,e^{\frac{1}{2}(w_2+w_3) tA}\dotsb e^{w_\nu tB}\dotsb e^{\frac{1}{2}w_1 tA} \label{eq:n2TypeSL}
  \end{align}
  with $w_i\in\mathbb{R}$. Note that, as shown in the last line, exponentials of $A$ for subsequent leapfrog terms can be contracted into one such that the total number of required operator exponentials is indeed $m$.
\end{itemize}

Depending on the number $m$ of factors in a decomposition $U$ of type \eqref{eq:n2TypeN}-\eqref{eq:n2TypeSL}, the parameters can be chosen such that $U$ coincides with the exact $e^{tH}$ up to order $p$ in the sense that
\begin{equation}\label{eq:constraint}
	U=e^{tH+\mc{O}(t^{p+1})}=e^{tH}+\mc{O}(t^{p+1}).
\end{equation}
Depending on $m$, the number of parameters in a decomposition can be larger than the number of constraints due to Eq.~\eqref{eq:constraint}. We can then use the remaining freedom to minimize the (leading-order) error term of the decomposition in a suitable metric.

Of course, for the same number of factors $m$, type SL is a subclass of type S, and type S is a subclass of type N,
\begin{equation}
	\tSL\subseteq\tS\subseteq\tN.
\end{equation}
At first sight, one might consider types S and SL to be superfluous. But we will see in Sec.~\ref{sec:n2NoParam} that types S and SL do not only have smaller numbers of parameters $\nu$, but also fewer constraints than a generic type-N decomposition for the same order $p$. For certain $m$ and $p$, type-N decompositions solving the constraint \eqref{eq:constraint} may then necessarily have to be of type S or such of type S may have to be of type SL. In this sense, below a certain minimum $m$ only decompositions of type S (or SL) may exist but none of type (SL or) N.

\subsection{Baker-Campbell-Hausdorff formula}\label{sec:BCH}
How can we compare a Lie-Trotter-Suzuki decomposition $U$ and $e^{tH}$ as suggested by Eq.~\eqref{eq:constraint}? One option would be to study the Taylor expansions of both operators in terms of operator monomials like $A^2BA$, require the difference of the expansions to vanish up to order $t^p$ and minimize some metric for the scalar coefficients of operator monomials in the $t^{p+1}$ term. However, in typical applications of many-body physics with finite-range interactions in $H$, norms for operator monomials of order $p+1$ scale as $\mc{O}(L^{p+1})$ with the system size $L$ which results in very large error bounds (cf.\ Sec.~\ref{sec:n2ErrorMeasure}).

We can use the Baker-Campbell-Hausdorff (BCH) formula \cite{Campbell1897-29,Baker1905-s2,Hausdorff1906,Dynkin1947-57,Dynkin1949,Varadarajan1984}
\begin{equation}\label{eq:BCH1}
	 \log\left(e^{X}e^{Y}\right)= X + Y + \frac{1}{2}[X,Y] + \frac{1}{12}[X,[X,Y]] - \frac{1}{12}[Y,[X,Y]] - \frac{1}{24}[Y,[X,[X,Y]]] \dots
\end{equation}
to resolve this issue. In particular, we can apply it recursively, to expand $\log U$ for the Lie-Trotter-Suzuki decompositions in terms of nested commutators of $A$ and $B$. Then, the constraints \eqref{eq:constraint} are equivalent to
\begin{equation}\label{eq:constraint2}
	Z=tH+\mc{O}(t^{p+1}) \quad \text{with} \quad Z:=\log U.
\end{equation}
In applications with finite-range interactions in $H$, norms (norm bounds) for the nested commutators in $Z$ will all scale \emph{linearly}  with the system size $L$.

To obtain $Z=\log U$ for a decomposition $U=e^{tA}\,e^{tB}\,e^{tC}\,e^{tD}\dotsb$, one can first use the BCH formula with
\begin{subequations}\label{eq:BCHiter}
\begin{equation}
	X_1=tA\quad \text{and}\quad Y_1=tB\quad \text{to get}\quad Z_1=\log(e^{tA}\,e^{tB}).
\end{equation}
Then, we apply it again with
\begin{equation}
	X_1=Z_1\quad \text{and}\quad Y_2=tC\quad \text{to get}\quad Z_2=\log(e^{Z_1}\,e^{tC})=\log(e^{tA}\,e^{tB}\,e^{tC})
\end{equation}
\end{subequations}
and so on until we arrive at $Z=\log U$.

\subsection{Free Lie algebra and Hall bases}\label{sec:n2HallBases}
Campbell, Baker, and Hausdorff \cite{Campbell1897-29,Baker1905-s2,Hausdorff1906} found that $\log\left(e^{X}e^{Y}\right)$ can be expressed in terms of nested commutators, i.e, that it is an element of the Lie algebra generated by $X$ and $Y$ \cite{Rossmann2002}. Dynkin finally derived an explicit formula \cite{Dynkin1947-57,Dynkin1949}.

The iteration \eqref{eq:BCHiter} of the BCH formula gives an expansion of $Z=\log U$ in terms of nested commutators of $A$ and $B$.
The number of constraints, imposed by requiring $U$ to coincide with $e^{tH}$ up to order $t^{p}$, can be determined from the number of terms in the expansion. In general, the nested commutators are not linearly independent, especially, because of the Jacobi identity
\begin{equation}\label{eq:JacobiIdenity}
	[V,[W,X]] + [W,[X,V]] + [X,[V,W]] = 0.
\end{equation}
\begin{table}[t]
	\setlength{\tabcolsep}{1.5ex}
	\begin{tabular}{c | l}
	 Degree & Hall basis elements\\
	 \hline
	 1	& $A$,\, $B$\\
	 2	& $[A,B]$\\
	 3	& $[A,[A,B]]$,\, $[B,[A,B]]$\\
	 4	& $[A,[A,[A,B]]]$,\, $[B,[A,[A,B]]]$,\, $[B,[B,[A,B]]]$\\
	 5	& $[A,[A,[A,[A,B]]]]$,\, $[B,[A,[A,[A,B]]]]$,\, $[B,[B,[A,[A,B]]]]$,\\
	 	& $[B,[B,[B,[A,B]]]]$,\, $[[A,B],[A,[A,B]]]$,\, $[[A,B],[B,[A,B]]]$\\
	 \hline
	 1	& $A$,\, $B$,\, $C$\\
	 2	& $[A,B]$,\, $[A,C]$,\, $[B,C]$\\
	 3	& $[A,[A,B]]$,\, $[A,[A,C]]$,\, $[B,[A,B]]$,\, $[B,[A,C]]$,\\
	 	& $[B,[B,C]]$,\, $[C,[A,B]]$,\, $[C,[A,C]]$,\, $[C,[B,C]]$\\
	 4	& $[A,[A,[A,B]]]$,\, $[A,[A,[A,C]]]$,\, $[B,[A,[A,B]]]$,\, $[B,[A,[A,C]]]$,\, $[B,[B,[A,B]]]$,\\
	 	& $[B,[B,[A,C]]]$,\, $[B,[B,[B,C]]]$,\, $[C,[A,[A,B]]]$,\, $[C,[A,[A,C]]]$,\, $[C,[B,[A,B]]]$,\\
	 	& $[C,[B,[A,C]]]$,\, $[C,[B,[B,C]]]$,\, $[C,[C,[A,B]]]$,\, $[C,[C,[A,C]]]$,\, $[C,[C,[B,C]]]$,\\
	 	& $[[A,B],[A,C]]$,\, $[[A,B],[B,C]]$,\, $[[A,C],[B,C]]$
	\end{tabular}
	\caption{\label{tab:HallBasis}Hall bases for the free Lie algebra generated by $n=2$ and $3$ operators, respectively.}
\end{table}
\begin{table}[t]
	\setlength{\tabcolsep}{1.5ex}
	\begin{tabular}{c | c c c c c c c c c c c c}
	 Degree $k$  & 1 & 2 & 3 & 4 & 5 & 6 & 7 & 8 & 9 & 10 & 11 & 12\\
	 \hline
	 $n=2$      & 2 & 1 & 2 & 3 & 6 & 9 & 18 & 30 & 56 & 99 & 186 & 335\\
	 $n=3$      & 3 & 3 & 8 & 18 & 48 & 116 & 312 & 810 & 2184 & 5880 & 16104 & 44220
	\end{tabular}
	\caption{\label{tab:HallBasisDim}Numbers $d_k$ of Hall bases elements with degree $k$ for the free Lie algebra generated by $n=2$ and $3$ operators, respectively.}
\end{table}

Let us discuss this problem more generally for $n$ operators instead of just two ($A$ and $B$). To resolve the above problem, we need a basis for the Lie algebra generated by operators $A,B,C,\dots$. For a \emph{free} Lie algebra, we can use Hall bases \cite{Hall1950-1,Serre1992,Reutenauer1993}.
A free Lie algebra is fully characterized by the properties of the commutator from which all algebraic relations like the Jacobi identity \eqref{eq:JacobiIdenity} follow and no further relations exit. For a finite vector space, the Lie algebra generated by $n$ operators will of course close at some point. However, for the large Hilbert spaces relevant in many-body physics and low expansion orders considered for our optimized decompositions, the assumption of a free Lie algebra is generally sufficient as we will see below.

For the construction of a Hall basis, one introduces a total order ``$<$'' on the generators and nested commutators, e.g.: $A<B<C<\dots$ for the generators, $X<Y$ for two nested commutators if $X$ is of lower degree than $Y$ and $[X,Y]<[V,W]$ if $X<V$, or $X=V$ and $Y<W$. According to Witt's formula \cite{Witt1956-64,Reutenauer1993}, the number of Hall basis elements of degree $k$ is given by the necklace polynomial
\begin{equation}\label{eq:HallBasisDim}
	d_k=\frac {1}{j}\sum _{j|k}\mu(j)\cdot n^{k/j},
\end{equation}
where $\mu$ denotes the M\"{o}bius function and the sum is over all integers $j$ that divide $k$. For $n=2$ and $3$ generators, Tables~\ref{tab:HallBasis} and \ref{tab:HallBasisDim} list Hall basis elements and their numbers $d_k$.
Note that Witt's formula \eqref{eq:HallBasisDim} addresses the case where all generators have degree one. For the decomposition type SL [Eq.~\eqref{eq:n2TypeSL}] and further decomposition for $n=3$ terms in Sec.~\ref{sec:n3Decomps}, we will also consider generators of higher degree.

\subsection{Numbers of parameters and constraints, symmetries}\label{sec:n2NoParam}
Assuming that all terms that can occur do occur in the expansion of $Z=\log U$ for an order-$p$ Lie-Trotter-Suzuki decomposition $U$, we can determine the number of constraints from the number of relevant Hall basis elements. For example, a type-N decomposition \eqref{eq:n2TypeN} with $m$ factors has $m$ parameters $\{a_i,b_i\}$ and the constraints due to Eq.~\eqref{eq:constraint2} are the polynomials in $\{a_i,b_i\}$ that appear in $Z$ as coefficients of Hall basis elements with degree $\leq p$. The numbers $\sum_{k=1}^p d_k$ of these are given in Eq.~\eqref{eq:HallBasisDim} and Table~\ref{tab:HallBasisDim}.

After taking into account symmetries etc., we will find that indeed all relevant Hall basis elements occur for the considered decompositions and expansion orders except for some obvious cases that are due to the fact that type SL decompositions are a subclass of those of type S, and the latter a subclass of the type N decompositions. To check and determine how many free parameters we actually have, after the expansion in a Hall basis, we can employ the Gr\"{o}bner basis of the constraint polynomials as discussed in the following section (Sec.~\ref{sec:Groebner}).

\textbf{Type N.} -- For the decompositions $U_{\tN,m}$ in Eq.~\eqref{eq:n2TypeN}, we have $\nu=m$ parameters $\{a_i,b_i\}$. As the decomposition has no further symmetry or structure, all Hall basis elements should occur in the expansion of $Z$ and the number of constraints to achieve approximation order $p$ is $\sum_{k=1}^p d_k$. For example, at first order, we have the two constraints $\sum_i a_i=\sum_i b_i = 1$ to achieve $Z=tH+\mc{O}(t^2)$.
\begin{table}[t]
	\setlength{\tabcolsep}{1.5ex}
	\begin{tabular}{c | l}
	 Degree & Hall basis elements\\
	 \hline
	 1	& $A_1$\\
	 3	& $A_3$\\
	 5	& $A_5$,\, $[A_1,[A_1,A_3]]$\\
	 7	& $A_7$,\, $[A_1,[A_1,A_5]]$,\, $[A_3,[A_1,A_3]]$,\, $[A_1,[A_1,[A_1,[A_1,A_3]]]]$\\
	 9	& $A_9$,\, $[A_1,[A_1,A_7]]$,\, $[A_3,[A_1,A_5]]$,\, $[A_5,[A_1,A_3]]$,\, $[A_1,[A_1,[A_1,[A_1,A_5]]]]$,\\
	 	& $[A_3,[A_1,[A_1,[A_1,A_3]]]]$,\, $[[A_1,A_3],[A_1,[A_1,A_3]]]$,\, $[A_1,[A_1,[A_1,[A_1,[A_1,[A_1,A_3]]]]]]$
	\end{tabular}
	\caption{\label{tab:HallBasisL}Hall basis elements of odd degree for the free Lie algebra generated by operators $\{A_1,A_3,A_5,\dotsc\}$, where $A_k$ has degree $k$.}
\end{table}
\begin{table}[t]
	\setlength{\tabcolsep}{1.5ex}
	\begin{tabular}{c | c c c c c c c c c c c c}
	 Degree $k$  & 1 & 3 & 5 & 7 & 9 & 11 & 13\\
	 \hline
	 $d_{\tL,k}$        & 1 & 1 & 2 & 4 & 8  & 18 & 40
	\end{tabular}
	\caption{\label{tab:HallBasisLDim}Numbers $d_{\tL,k}$ of Hall bases elements with odd degree $k$ for the free Lie algebra generated by operators $\{A_1,A_3,A_5,\dotsc\}$, where $A_k$ has degree $k$.}
\end{table}

\textbf{Type S}. -- For the decompositions $U_{\tS,m}$ in Eq.~\eqref{eq:n2TypeS}, we have $\nu=(m+1)/2$ parameters and odd $m$. Due to the symmetry in the factors, the decompositions obey the time reversal symmetry $U_{\tS,m}(t)U_{\tS,m}(-t)=\id$. It follows that $Z(t)$ only contains terms of odd order in $t$ such that no Hall basis elements of even degree occur \cite{Yoshida1990}:
\begin{lemma}[\lemmaHead{Time reversal symmetry and order}]\label{lemma:n2ReversalSym}
  Let $U(t)$ be of the form $U(t)=e^{tA}\,e^{tB}\,e^{tC}\dotsc$ and obey time reversal symmetry $U(t)U(-t)=\id$. In an expansion $\log U(t)=tZ^{(1)}+t^2 Z^{(2)}+t^3 Z^{(3)}+t^4 Z^{(4)}+\dots$, all even-order terms vanish, i.e., $Z^{(2)}=Z^{(4)}=Z^{(6)}=\dots=0$, and $Z^{(1)}=A+B+C+\dots$
\end{lemma}
As an immediate consequence, a symmetric order $p$ decomposition (obeys $Z^{(1)}=H$ and $Z^{(2)}=Z^{(3)}=\dots=Z^{(p)}=0$) with odd $p$ is actually also an order $p+1$ decomposition \cite{Yoshida1990,Suzuki1990}. For the proof of Lemma~\ref{lemma:n2ReversalSym}, note that $\log U(-t)=-tZ^{(1)}+t^2 Z^{(2)}-t^3 Z^{(3)}+t^4 Z^{(4)}+\dots$ Applying the BCH formula \eqref{eq:BCH1}, we find $\log[U(t)U(-t)]=2t^2 Z^{(2)}+\mc{O}(t^3)$ such that $Z^{(2)}=0$ because of the constraint $U(t)U(-t)=\id$. With this information, we can reconsider the BCH formula and find $\log[U(t)U(-t)]=2t^4 Z^{(4)}+\mc{O}(t^5)$, showing that $Z^{(4)}=0$. Continuing in this way, we find that all even-order terms vanish.

For a symmetric decomposition that has no further structure, all Hall basis elements of odd degree should occur in the expansion of $Z$ and the number of constraints to achieve approximation order $p$ is $\sum_{q,2q-1\leq p} d_{2q-1}$. For example, at first order, we have the two constraints that the coefficients $a_i$ in the $A$ factors and the coefficients $b_i$ in the $B$ factors in Eq.~\eqref{eq:n2TypeS} sum to one to achieve $Z=tH+\mc{O}(t^2)$.
\begin{table}[t]
	\setlength{\tabcolsep}{1.5ex}
	Number $N^\tH_p$ of Hall basis elements with degree $\leq p$ in an expansion of $\log U$ with $n=2$:\\[0.5em]
	\begin{tabular}{c || c | c | c | c | c | c | c | c | c | c | c | c}
	 order $p$   & 1 & 2 & 3 & 4 & 5 & 6 & 7 & 8 & 9 & 10 & 11 & 12\\
	 \hline
 	 Type N  & 2 & 3 & 5 & 8 & 14 & 23 & 41 & 71 & 127 & 226 & 412 & 747\\
	 Type S  & --& 2 & --& 4 & -- & 10 & -- & 28 &  -- & 84  &  -- & 270\\
	 Type SL & --& 1 & --& 2 & -- & 4  & -- &  8 &  -- & 16  &  -- & 34
	\end{tabular}\\[1em]
	Minimum number $m_p^\tH$ of factors needed to allow for  $\log U= tH + \mc{O}(t^{p+1})$ with $n=2$:\\[0.5em]
	\begin{tabular}{c || c | c | c | c | c | c | c | c | c | c | c | c}
	 order $p$      & 1 & 2 & 3 & 4 & 5 & 6 & 7 & 8 & 9 & 10 & 11 & 12\\
	 \hline
 	 Type N  & 2 & 3 & 5 & 8 & 14 & 23 & 41 & 71 & 127 & 226 & 412 & 747\\
	 Type S  & --& 3 & --& 7 & -- & 19 & -- & 55 &  -- & 167 &  -- & 539\\
	 Type SL & --& 3 & --& 7 & -- & 15 & -- & 31 &  -- &  63 &  -- & 135
	\end{tabular}
	\caption{\label{tab:n2NoConstraints}For the decompositions $U$ defined in Sec.~\ref{sec:n2Decomps} with $n=2$ non-commuting terms in $H$, the first table gives the number $N^\tH_p$ of Hall basis elements with degree $\leq p$ in an expansion of $\log U$. Assuming that the resulting $N^\tH_p$ constraints to obtain an order-$p$ decomposition of $e^{tH}=e^{t(A+B)}$ are independent, the second table gives the corresponding minimum number $m_p^\tH$ of factors needed to for order-$p$ decompositions.}
\end{table}

\textbf{Type SL.} -- For the decompositions $U_{\tSL,m}$ in Eq.~\eqref{eq:n2TypeSL}, we have $\nu=\ceil((m-1)/4)$ parameters and odd $m$. As these decompositions are symmetric, according to Lemma~\ref{lemma:n2ReversalSym} only terms of odd degree occur in the expansion of $Z=\log U_{\tSL,m}$. The decomposition is a product of leapfrog terms $U_\tL(\tau)$ [Eq.~\eqref{eq:leapfrog}] which are symmetric second order decompositions of $e^{\tau H}$ and can hence be expanded in the form
\begin{equation}\label{eq:leapfrogTerms}
	\log U_\tL(\tau)= \tau Z^{(1)}_\tL + \tau^3 Z^{(3)}_\tL + \tau^5 Z^{(5)}_\tL+ \dots \quad\text{with}\quad
	Z^{(1)}_\tL=H.
\end{equation}
$Z^{(k)}_\tL$ is a Lie polynomial containing only nested commutators of degree $k$. To determine the number of constraints for $U_{\tSL,m}$ to be an order-$p$ decomposition of $e^{t H}$, we can consider the Lie algebra generated by $\{Z^{(1)}_\tL,Z^{(3)}_\tL,Z^{(5)}_\tL,\dots\}$. Again, assuming no further relevant algebraic relations for the generators, we can treat the problem as that of a free Lie algebra. The number of constraints due to Eq.~\eqref{eq:constraint2} is then given by the number $\sum_{q,2q-1\leq p} d_{\tL,2q-1}$ of Hall basis elements of odd degree $2q-1\leq p$. Here, for example, the degree of $[Z^{(1)}_\tL,[Z^{(1)}_\tL,Z^{(5)}_\tL]]$ is 7. Tables~\ref{tab:HallBasisL} and \ref{tab:HallBasisLDim} list the Hall basis elements and their numbers $d_{\tL,k}$.  At first order, we only have the constraint that the coefficients $w_i$ of all leapfrog factors in Eq.~\eqref{eq:n2TypeSL} sum to one to achieve $Z=tH+\mc{O}(t^2)$.

To summarize this section, we give the total numbers of constraints by order in Table~\ref{tab:n2NoConstraints}. The table also states $m_p^\tH$, the minimum number of factors needed in the different decompositions to obtain an order-$p$ decomposition [cf.\ Eq.~\eqref{eq:constraint}], under the assumption that all constraint polynomials are independent. This is of course not true in all cases. For example, for sixth order decompositions we have $m_p^\tH=15,19,23$ for types SL, S, and N, respectively. But still, any decomposition of type SL (S) is at the same time of type S (N). How to reconcile these statements? If we want to construct sixth order decompositions of type S or N with $m<19$, we can do so, but they turn out to be of the specific type SL, i.e., consist of leapfrog factors. And if we want to construct a type-N decomposition with $19\leq m<23$, we can do so, but they turn out to be of the specific type S, i.e., symmetric.

Although we have not encountered this situation in our current search for optimized decomposition, it could also occur, that the constraint polynomials (the coefficients of the different Hall basis elements discussed above) are not independent. For example, a polynomial could be a multiple of another one.

\subsection{Ideal of constraint polynomials and Gr\"{o}bner bases}\label{sec:Groebner}
For decompositions $U$ as defined in Sec.~\ref{sec:n2Decomps}, the constraints to obtain a decomposition of order $p$ [Eq.~\eqref{eq:constraint2}] are polynomials in the $\nu$ parameters of the decomposition ($\{a_i,b_i\}$ or $\{w_i\}$) and can be read off as the coefficients of Hall basis elements in an expansion of $Z=\log U$. If the constraints are all independent, the number of free parameters is simply given by $\nu-N^\tH_p$. Here, $N^\tH_p$ denotes the number of Hall basis elements with degree $\leq p$ in an expansion of $Z$ (cf.\ Table~\ref{tab:n2NoConstraints}).

The independence of the constraints can be checked by constructing a Gr\"{o}bner basis for the ideal defined by the constraint polynomials \footnote{For a finite set $\mc{C}=\{c_1,\cdots, c_k\}$ of polynomials for variables $\vec{x}$, the ideal generated by $\mc{C}$ is the set of linear combinations of the $c_i$ with arbitrary polynomial coefficients.}. The Gr\"{o}bner basis \cite{Buchberger1965,Adams1994} -- a generating set for the ideal -- can also be used to determine maximal sets of independent variables or the number of solutions if it is finite. Finally, the Gr\"{o}bner basis can be used to find solutions for the system of polynomial equations. Algorithms to compute Gr\"{o}bner bases were given by Buchberger \cite{Buchberger1976-10,Cox2007} and Faug\`{e}re \cite{Faugere1999-139,Faugere2002}.

\subsection{Relevance of 1-norm as an error measure}\label{sec:n2ErrorMeasure}
Above, we have discussed how to construct Lie-Trotter-Suzuki decompositions $U$ as products of operator exponentials to approximate $e^{tH}=e^{t(A+B)}$ to order $p$ [Eq.~\eqref{eq:constraint}]. The more factors we include in a decomposition, the more free parameters we have which allow us to reduce the amplitude of deviations from $e^{tH}$. We will only consider the leading error term which is of order $t^{p+1}$. Like McLachlan in Ref.~\cite{McLachlan1995} we will use the free parameters in the decomposition to minimize a measure for the leading error term.

We can expand the leading error term in elements $\{X_i\}$ of degree $p+1$ from a Hall basis that is generated by $A$ and $B$ such that
\begin{equation}\label{eq:Uexpansion}\ts
	\log U = tH + t^{p+1}\sum_i c_i X_i + \mc{O}(t^{p+2})
\end{equation}
The relevant error measure is the operator norm distance which we can bound using the triangle inequality such that
\begin{subequations}\label{eq:normDiff}
\begin{equation}\ts
	\|\log U - tH\|\, =\, t^{p+1}\|\sum_i c_i X_i\| + \mc{O}(t^{p+2}) 
	\,\leq\, t^{p+1}\sum_i |c_i|\, \|X_i\| + \mc{O}(t^{p+2}).
\end{equation}
Equivalently, we have of course
\begin{equation}\ts
	U=e^{tH}+ t^{p+1}\sum_i c_i X_i + \mc{O}(t^{p+2})\quad\text{and}\quad
	\|U-e^{tH}\| \leq t^{p+1}\sum_i |c_i|\, \|X_i\| + \mc{O}(t^{p+2}).
\end{equation}
\end{subequations}

The $X_i$ are nested commutators of $A$ and $B$ with degree $p+1$. In our generic optimization we do not want to assume any detailed information about $A$ and $B$. To simplify matters further, we use a uniform norm bound for all $X_i$: For a typical situation where $A$ and $B$ are both sums of finite-range interaction terms with disjoint spatial supports, the number of terms in all $X_i$ will be linear in the system size $L$. However, the spatial support of a term $[A,[A,[A,[A,B]]]]$ would then be smaller than that of terms like $[B,[A,[B,[A,B]]]]$ and one could use these properties to further improve the error measure.

So, with a uniform norm bound on the $X_i$ we can use the 1-norm $\sum_i |c_i|$ of the expansion coefficients as an error measure. There is one more complication. The $c_i$ can and often do depend on the choice of the Hall basis. The latter depends on the order that is chosen for the generators $A$ and $B$. As both $A<B$ and $B<A$ are perfectly fine and result in a valid upper bound on the norm distance, we can minimize with respect to the ordering. Let the two corresponding sets of coefficient polynomials be denoted by $\{c^{AB}_i\}$ and $\{c^{BA}_i\}$, respectively. We will then use the error measure
\begin{equation}\label{eq:n2Error}\ts
	\veps := \left(\frac{m}{p}\right)^p\min\left( \sum_i |c^{AB}_i|,\, \sum_i |c^{BA}_i|\right)
\end{equation}
to quantify the magnitude of the deviation of $U$ from $e^{tH}$, where $m$ denotes again the number of factors in the decomposition $U$. Like the coefficient polynomials $c_i$, the error $\veps$ is a function of the parameters in the decomposition.

Similar to Ref.~\cite{McLachlan1995}, the prefactor $m^p$ in Eq.~\eqref{eq:n2Error} has been chosen to allow for a fair comparison of decompositions with the same order $p$, but different numbers $m_j$ of factors: We want to compare the accuracies for evolving the system for a time $T$ at constant computation cost. The evolution can be accomplished by applying the decomposition $U$ with time step size $t$ for $T/t$ times. Consider two order-$p$ decompositions with $m_1$ and $m_2$ factors, respectively. Assuming that the computational cost for the implementation of the exponentials $e^{at A}$ and $e^{bt B}$ is uniform or comparable, we should choose time steps of size $t_j\propto m_j$. This scaling of the time step, the prefactor $t^{p+1}$ of the leading error term in Eq.~\eqref{eq:normDiff}, and the number $T/t$ of time steps motivate the factor $m^p$ in Eq.~\eqref{eq:n2Error}. The additional factor $1/p^p$ is irrelevant and just added in order to prohibit $\veps$ from increasing too much when increasing $p$. Note that $\veps$ can \emph{not} be used to compare decompositions of different order $p$.

We have discussed above that norm bounds for the nested commutators in Eqs.~\eqref{eq:normDiff} are linear in the system size $L$ for the case of finite-range interactions. In fact this is overpessimistic in many situations, in particular, if we are only interested in the evolution of local quantities, i.e., observables which are a sum of operators with finite spatial support. For the unitary time evolution of closed systems and Markovian dynamics of open quantum systems, $L$ can be replaced by $(vT)^d$, where $d$ denotes the number of spatial dimensions, $T$ the maximum time for which we want to evolve, and $v$ is a Lieb-Robinson velocity \cite{Lieb1972-28,Poulin2010-104,Nachtergaele2011-552,Barthel2012-108b}. This is due to \emph{quasi-locality} \cite{Nachtergaele2007-12a,Barthel2012-108b}: In the Heisenberg picture, one can truncate the evolution outside a region of size $\sim(vT)^d$ around the spatial support of a given local observable.

\subsection{Unoptimized decompositions due to Forest, Ruth, Yoshida, and Suzuki}\label{sec:n2Unoptimized}
Based on the symmetric second order (leapfrog) decomposition $U_\tL$ in Eq.~\eqref{eq:leapfrog}, Forest and Ruth \cite{Forest1990-43} found the fourth order decomposition
\begin{equation}
	U_\tL(wt)\,U_\tL((1-2w)t)\,U_\tL(wt)\,=\,e^{tH}+\mc{O}(t^{5})
	\quad\text{with}\quad
	w=(2-2^{1/3})^{-1}\approx 1.35121
\end{equation}
consisting of three leapfrog factors. It is hence of type SL with $m=7$ in Eq.~\eqref{eq:n2TypeSL}.

Yoshida \cite{Yoshida1990} generalized the approach. Let us define $U_{\tY,1}(t):=U_\tL(t)$ for the leapfrog decomposition \eqref{eq:leapfrog}. Yoshida then showed how to obtain decompositions $U_{\tY,q}$ of arbitrary (even) order $p=2q$ through the recursion
\begin{equation}\label{eq:Yoshida}
\begin{split}
	&U_{\tY,q+1}(t) := U_{\tY,q}(y_q t)\, U_{\tY,q}((1-2y_q)t)\, U_{\tY,q}(y_q t) \,=\,e^{tH}+\mc{O}(t^{2q+3})\\
	&\text{with}\quad y_q=\left(2-2^{1/(2q+1)}\right)^{-1}.
\end{split}
\end{equation}
This approach has two drawbacks: (a) the number $m$ of factors in the decompositions is $2\cdot 3^{q-1}+1 = 3,7,19,55,163,487,1459,\dotsc$ and grows considerably faster than the theoretical minimum $m_p^\tH$ for type-SL decompositions given in Table~\ref{tab:n2NoConstraints}, and (b) the parameters $y_q=1.35121,\,1.17467,\,1.11618,\dotsc$ are all larger than one. This results in large error values $\veps$ as we will see below and the series of decompositions does not converge in the limit $q\to\infty$.

Suzuki \cite{Suzuki1990} resolved the latter issue with what he called fractal decompositions, using more factors than above. Let us define $U_{\tZ,1}(t):=U_\tL(t)$ for the leapfrog decomposition. Suzuki then showed how to obtain decompositions $U_{\tZ,q}$ of arbitrary (even) order $p=2q$ through the recursion
\begin{equation}\label{eq:Suzuki}
\begin{split}
	&U_{\tZ,q+1}(t) := U_{\tZ,q}(z_q t)\, U_{\tZ,q}(z_q t)\, U_{\tZ,q}((1-4z_q)t)\, U_{\tZ,q}(z_q t)\, U_{\tZ,q}(z_q t)
	 \,=\,e^{tH}+\mc{O}(t^{2q+3})\\
	&\text{with}\quad z_q=\left(4-4^{1/(2q+1)}\right)^{-1}.
\end{split}
\end{equation}
Now, both $z_q=0.414491,\,0.373066,\,0.359585,\,\dotsc$ and $|1-4z_q|$ are smaller than one. Specifically, $\frac{1}{3}<z_q<\frac{1}{2}$ and $|1-4z_q|<\frac{2}{3}$ \cite{Suzuki1990}. Still, this approach has drawbacks: (a) the number $m$ of factors in the decompositions is $2\cdot 5^{q-1}+1 = 3,11,51,251,1251,6251,31251\dotsc$ and grows hence even faster than in Yoshida's scheme, and (b) the resulting error values $\veps$ are still much larger than those of the optimized decompositions discussed below.

\section{Optimized decompositions for \texorpdfstring{$n=2$}{n=2} terms}\label{sec:n2opt}
Here, we optimize the decompositions of types N, S, SL defined in Sec.~\ref{sec:n2Decomps} with respect to their parameters to minimize the error measure $\veps$ in Eq.~\eqref{eq:n2Error}. Keep in mind, that $\veps$ is properly scaled to allow for a fair comparison of decompositions with the same order $p$ but different numbers of factors $m$; it takes into account that, for larger $m$, also the time step $t$ should be increased to compare integrators with the same computation costs.

We compare the results to the unoptimized type-SL decompositions of Forest and Ruth  \cite{Forest1990-43}, Yoshida \cite{Yoshida1990}, and Suzuki \cite{Suzuki1990}. In most cases, we find that the optimal decompositions are relatively close to those obtained by McLachlan \cite{McLachlan1995} who used a different error measure. He expanded in a somewhat different basis suitable for symplectic integrators in classical Hamiltonian systems and, for practical reasons, used the 2-norm of the expansion coefficients instead of the more relevant 1-norm \eqref{eq:n2Error}. The definition for McLachlan's ``Hamiltonian truncation error'' is given in Ref.~\cite{McLachlan1992-5}.

In the discussion below, $\nu$ denotes the number of parameters as stated in Sec.~\ref{sec:n2Decomps} and the number of constraints for the different decomposition types and approximation orders $p$ are given in Table~\ref{tab:n2NoConstraints}. We check the applicability of the corresponding counting argument using Gr\"{o}bner bases for the constraint polynomials as discussed in Sec.~\ref{sec:Groebner}. For each order $p$, the recommended decomposition (usually smallest $\veps$ found) is indicated by a star. When the two possible orders $A<B$ and $B<A$ chosen for the construction of the Hall basis are not equivalent, we specify in brackets the order that yields the minimum for $\veps$.

\subsection{Order \texorpdfstring{$p=2$}{p=2}}
\myitem
\emph{Leapfrog ($m=3$, type SL, $\nu=1$).} -- The only parameter is fixed to $w_1=1$ by the constraint that the first order term in $\log U$ is $tH=t(A+B)$. The error is $\veps = \left(\frac{3}{2}\right)^2\frac{1}{8} = 0.28125$.

\myitem
\emph{McLachlan ($m=5$, type S, $\nu=3$).} -- There are two constraints and hence one free parameter. McLachlan states
\begin{equation}\ts
	b_1=\frac{1}{2},\quad a_2=1-2a_1,\quad a_1=\frac{1}{12y}(y^2+6y-2)\approx 0.19318 \quad\text{with}\quad
	y=(2\sqrt{326}-36)^{1/3}
\end{equation}
which has error $\veps\approx 0.075192$.

\myitemBest
\emph{Optimized ($m=5$, type S, $\nu=3$).} -- There are two constraints and hence one free parameter; we choose $a_1$. The error is minimized for
\begin{equation}\label{eq:dec-n2p2Best}\ts
	b_1=\frac{1}{2},\quad a_2=1-2a_1,\quad a_1=\frac{1}{6}(3-\sqrt{3})\approx 0.21132
\end{equation}
which gives $\veps\approx 0.069778$.  At this point, the coefficient of the term $[A,[A,B]]$ vanishes.

\myitemSummary
\emph{Discussion.} -- The best decomposition found here is of type S with $m=5$ as specified in Eq.~\eqref{eq:dec-n2p2Best}. It improves over the common leapfrog decomposition by a factor $\sim 1/4$. According to Table~\ref{tab:n2NoConstraints}, we can reach order $p=4$ with $m=7$ factors.

\subsection{Order \texorpdfstring{$p=4$}{p=4}}
\myitem
\emph{Forest \& Ruth, Yoshida ($m=7$, type SL).} -- For the decomposition $U_{\tY,q=2}$ with $q=2$ in Eq.~\eqref{eq:Yoshida}, the error is $\veps \approx 0.38640$ ($A<B$).

\myitem
\emph{McLachlan, Omelyan et al.\ ($m=9$, type S, $\nu=5$).} -- There are four constraints and hence there is one free parameter. McLachlan states
\begin{equation}\label{eq:dec-n2p4m9S-McL}
\begin{split}
	&\ts b_2=\frac{1}{2}-b_1,\quad a_3=1-2(a_1+a_2),\quad b_1=\frac{6}{11}\approx 0.54545,\\
	&\ts a_1 = \frac{1}{3924}\left(642+\sqrt{471}\right)\approx 0.16914,\quad a_2 = \frac{121}{3924}\left(12-\sqrt{471}\right)\approx -0.29919
\end{split}
\end{equation}
which has error $\veps\approx 0.072483$ ($B<A$). Optimizing the coefficient 2-norm, Omelyan et al.\ \cite{Omelyan2002-146} find
\begin{equation}\label{eq:dec-n2p4m9S-Omel}
	a_1 = 0.1720865590295143,\quad
	b_1 = 0.5915620307551568,\quad
	a_2 = -0.1616217622107222
\end{equation}
which gives $\veps\approx 0.069248$ ($B<A$).

\begin{figure}[p]
  \begin{tabular}{l@{\hspace{7ex}}l}
   {\small{(a)}}\quad \underline{$p=4$, $m=9$, type S}
  &{\small{(b)}}\quad \underline{$p=4$, $m=11$, type SL}\\[-0em]
                \imagetop{\includegraphics[height=0.24\columnwidth]{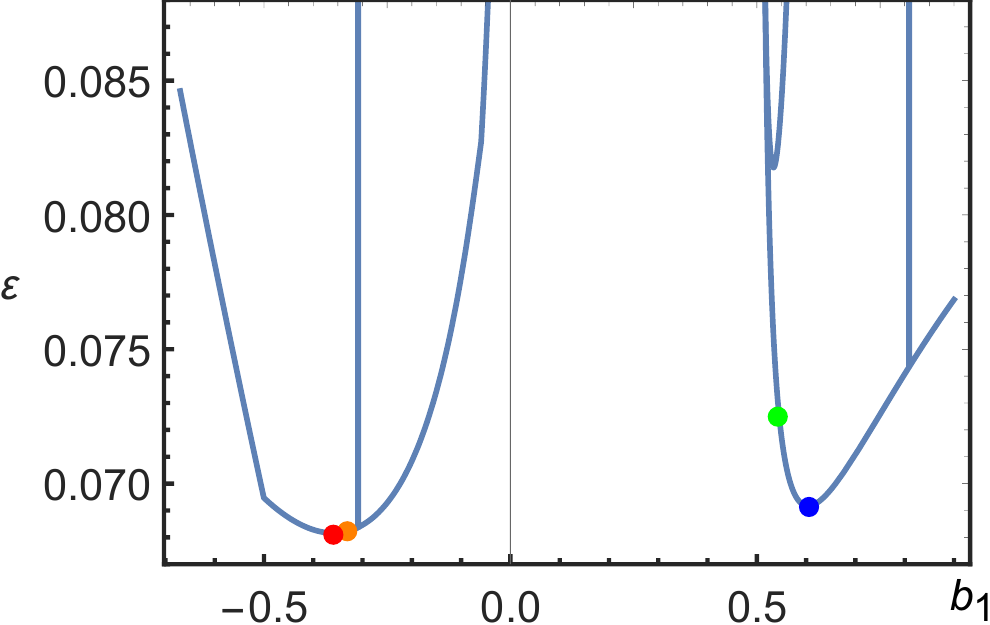}}\vspace{1.2em}
  &             \imagetop{\includegraphics[height=0.24\columnwidth]{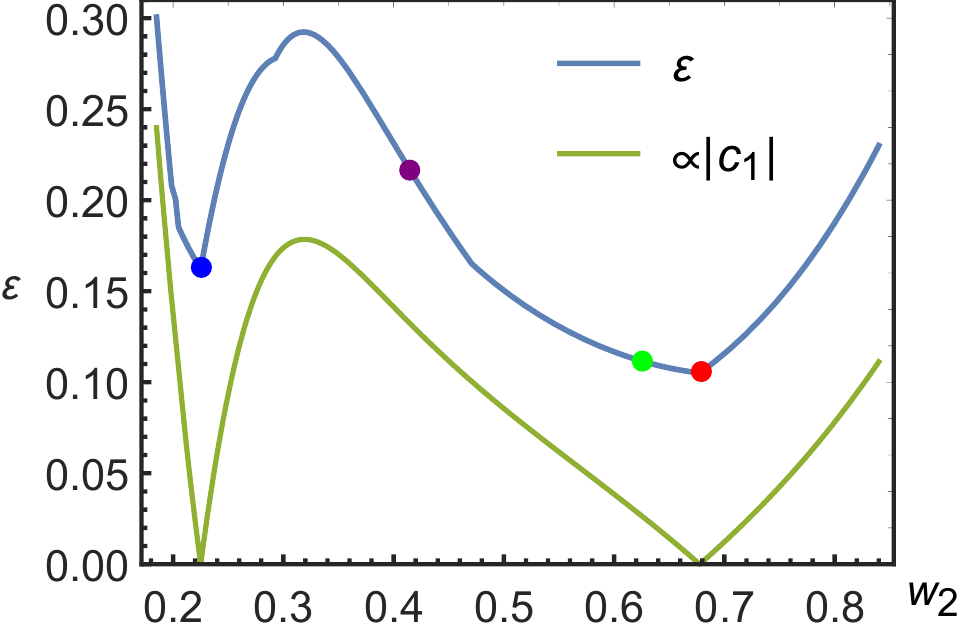}}\\[0.2em]
   {\small{(c)}}\quad \underline{$p=4$, $m=11$, type S}
  &{\small{(d)}}\quad \underline{$p=4$, $m=11$, type S}\\[-0em]
                \imagetop{\includegraphics[height=0.25\columnwidth]{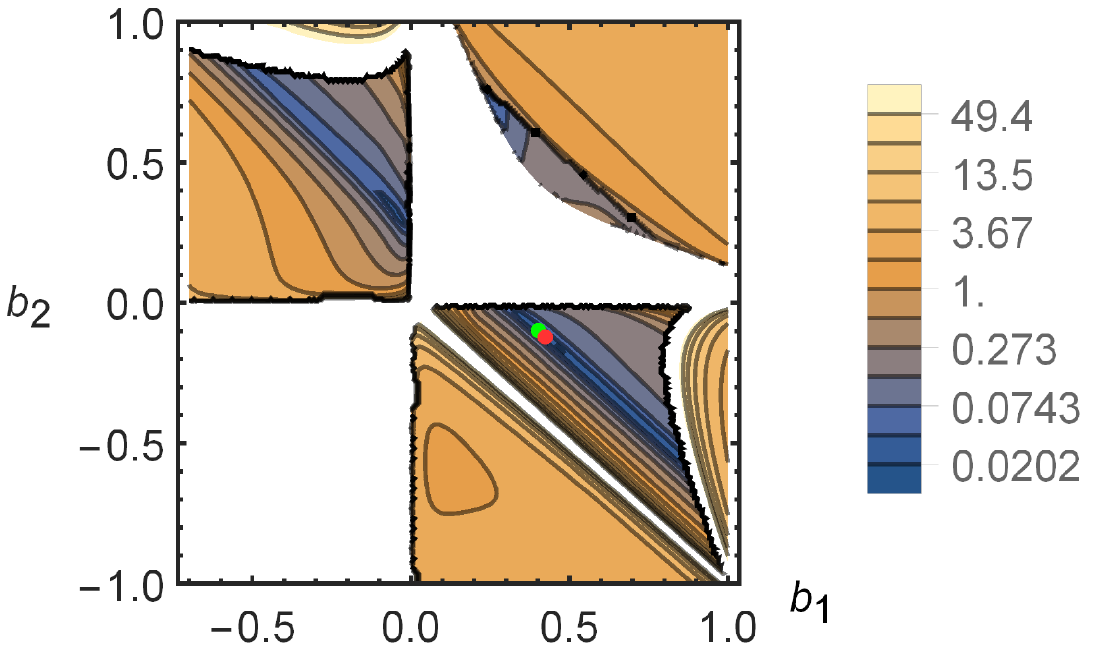}}\vspace{1.2em}
  &             \imagetop{\includegraphics[height=0.25\columnwidth]{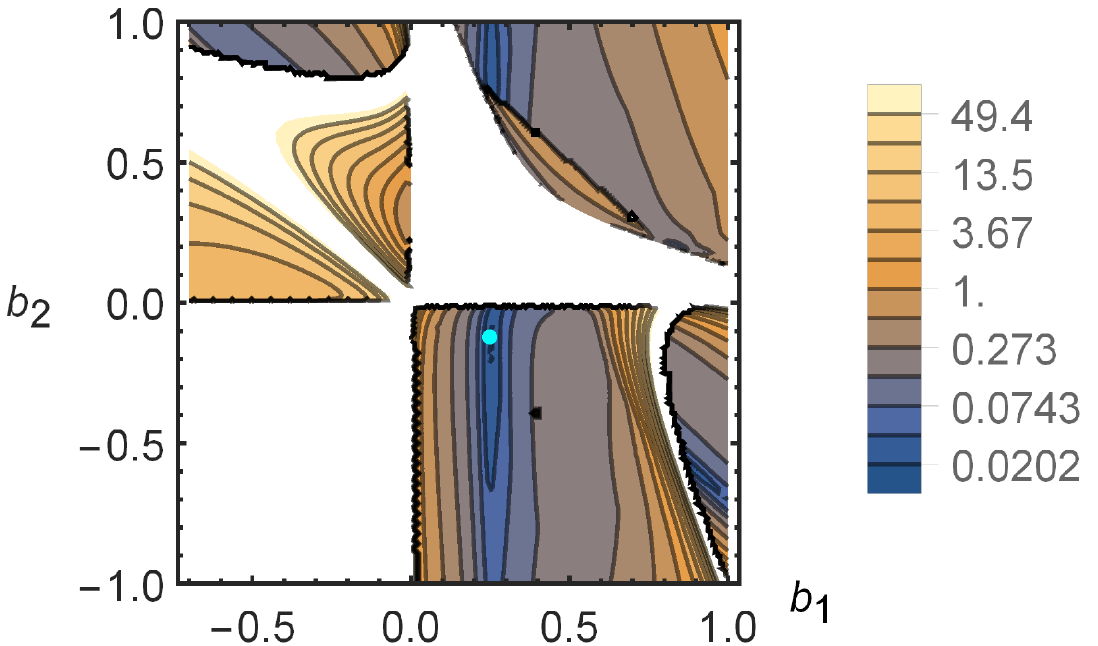}}\\[0.2em]
   {\small{(e)}}\quad \underline{$p=6$, $m=19$, type SL}
  &{\small{(f)}}\quad \underline{$p=6$, $m=23$, type SL}\\[-0em]
                \imagetop{\includegraphics[height=0.235\columnwidth]{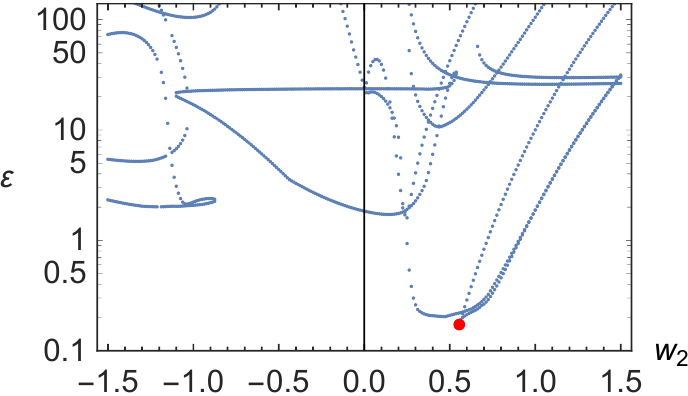}}
  &             \imagetop{\includegraphics[height=0.25\columnwidth]{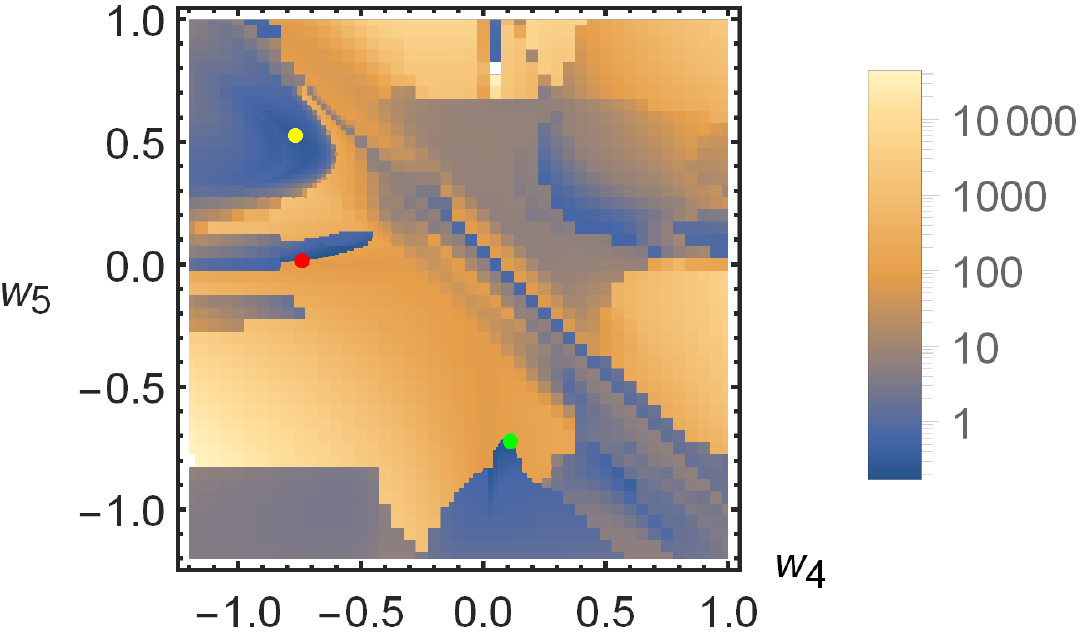}}
  \end{tabular}
  \caption{\label{fig:n2Errors}Error measures $\veps$ for various decompositions with $n=2$ terms and locations of minimal solutions compared to locations of solutions from the literature.  \textbf{(a)} Fourth order S decompositions with $m=9$ factors. Red: global minimum \eqref{eq:dec-n2p4m9S-1}, orange: nearby analytical solution \eqref{eq:dec-n2p4m9S-1A}, blue: second local minimum \eqref{eq:dec-n2p4m9S-2}, green: McLachlan's result \eqref{eq:dec-n2p4m9S-McL}.
  \textbf{(b)} Fourth order SL decompositions with $m=11$ factors. Red: global minimum \eqref{eq:dec-n2p4m11SL-1}, orange: nearby analytical solution \eqref{eq:dec-n2p4m11SL-1A}, blue: second local minimum \eqref{eq:dec-n2p4m11SL-2}, green: McLachlan's result \eqref{eq:dec-n2p4m11SL-McL}, purple: Suzuki's solution \eqref{eq:Suzuki}. At the minima, the coefficient $|c_1|$ of terms $[A,[A,[A,[A,B]]]]$ and $[B,[A,[A,[A,B]]]]$ in the expansion \eqref{eq:Uexpansion} vanishes.
  \textbf{(c,d)} Fourth order S decompositions with $m=11$ factors. Red: global minimum \eqref{eq:dec-n2p4m11S-1}, green: McLachlan's result \eqref{eq:dec-n2p4m11S-McL}, cyan: second local minimum \eqref{eq:dec-n2p4m11S-2}.
  \textbf{(e)} Sixth order SL decompositions with $m=19$ factors. Red: global minimum \eqref{eq:dec-n2p6m19SL}.
  \textbf{(f)} Sixth order SL decompositions with $m=23$ factors. Red: global minimum \eqref{eq:dec-n2p6m23SL}, green: second local minimum with $\veps\approx 0.19599$, yellow: third local minimum with $\veps\approx 0.26551$.
  }
\end{figure}

\myitem
\emph{Optimized ($m=9$, type S, $\nu=5$).} -- There are four constraints and hence there is one free parameter; according to the Gr\"{o}bner basis, we can choose $b_1$. The error is minimized for
\begin{subequations}
\begin{equation}\label{eq:dec-n2p4m9S-1}
\begin{split}
	&\ts b_2=\frac{1}{2}-b_1,\quad a_3=1-2(a_1+a_2),\quad b_1=-0.35905925216967795307, \\
	&\ts a_1 = 0.26756486526206148829,\quad a_2 =-0.034180403245134195595
\end{split}
\end{equation}
which gives $\veps\approx 0.068161$ ($B<A$). A nearby analytical solution with almost identical error is
\begin{equation}\label{eq:dec-n2p4m9S-1A}
\begin{split}
	&\ts b_2=\frac{1}{2}-b_1,\quad a_3=1-2(a_1+a_2),\quad b_1=-\frac{1}{3},\\
	&\ts a_1 = \frac{17}{2}-\frac{5}{2} \sqrt{\frac{65}{6}},\quad a_2 =\frac{3}{20} \left(\sqrt{390}-20\right).
\end{split}
\end{equation}
There is another local minimum at
\begin{equation}\label{eq:dec-n2p4m9S-2}
\begin{split}
	&\ts b_2=\frac{1}{2}-b_1,\quad a_3=1-2(a_1+a_2),\quad b_1=0.60417497648530223585, \\
	&\ts a_1 = 0.17285948240376668244,\quad a_2 =-0.14265971252922336963.
\end{split}
\end{equation}
It has error $\veps\approx 0.069172$ ($B<A$) and is relatively close to the results in Eqs.\ \eqref{eq:dec-n2p4m9S-McL} and \eqref{eq:dec-n2p4m9S-Omel}. See Fig.~\ref{fig:n2Errors}(a).
\end{subequations}

\myitem
\emph{Suzuki, Kahan \& Li ($m=11$, type SL).} -- For Suzuki's decomposition $U_{\tZ,q=2}$ in Eq.~\eqref{eq:Suzuki}, the error is $\veps\approx  0.216883$ ($A<B$). In Ref.~\cite{Kahan1997-66}, Kahan and Li state the two solutions
\begin{equation}\label{eq:dec-n2p4m11SL-KahanLi}\ts
	w_3=1-2(w_1+w_2),\quad w_1=\frac{3\pm\sqrt{3}}{6},\quad w_2 = \frac{3\mp\sqrt{3}}{6}
\end{equation}
which have the error $\veps\approx 0.17706$ ($A<B$).

\myitem
\emph{McLachlan, Omelyan et al.\ ($m=11$, type SL, $\nu=3$).} -- There are two constraints and hence there is one free parameter. McLachlan states
\begin{equation}\label{eq:dec-n2p4m11SL-McL}
	w_3=1-2(w_1+w_2),\quad w_1=0.28,\quad w_2 = 0.62546642846767004501
\end{equation}
which has error $\veps\approx 0.11155$ ($A<B$). Optimizing the coefficient 2-norm, Omelyan et al.\ \cite{Omelyan2002-146} find
\begin{equation}\label{eq:dec-n2p4m11SL-Omel}
	w_3=1-2(w_1+w_2),\quad w_1=0.3221375960817984,\quad w_2 = 0.5413165481700430
\end{equation}
which gives $\veps\approx 0.13365$ ($A<B$).

\myitem
\emph{Optimized ($m=11$, type SL, $\nu=3$).} -- There are two constraints and hence there is one free parameter; according to the Gr\"{o}bner basis, we can choose $w_2$. The error is minimized for
\begin{subequations}
\begin{equation}\label{eq:dec-n2p4m11SL-1}
	w_3=1-2(w_1+w_2),\quad w_1=0.25686635900587695859,\quad w_2 =0.67762403230558747362
\end{equation}
which gives $\veps\approx 0.10509$ ($A<B$). At this point, the coefficients of the terms $[A,[A,[A,[A,B]]]]$ and $[B,[A,[A,[A,B]]]]$ in the expansion \eqref{eq:Uexpansion} vanish. A nearby analytical solution with almost identical error is
\begin{equation}\label{eq:dec-n2p4m11SL-1A}\ts
	w_3=1-2(w_1+w_2),\ \ w_1=\frac{1}{18} \left(\left(278-6 \sqrt{2145}\right)^{1/3}+\left(278+6\sqrt{2145}\right)^{1/3}-4\right),\ \ w_2 =\frac{2}{3}.
\end{equation}
Another local minimum where the same error coefficients vanish is located at
\begin{equation}\label{eq:dec-n2p4m11SL-2}
	w_3=1-2(w_1+w_2),\quad w_1=0.75433412633084310590,\quad w_2 =0.22503541239785228348.
\end{equation}
It has error $\veps\approx 0.16224$ ($A<B$). See Fig.~\ref{fig:n2Errors}(b).
\end{subequations}

\myitem
\emph{McLachlan ($m=11$, type S, $\nu=6$).} -- There are four constraints and hence two free parameters. McLachlan states
\begin{equation}\label{eq:dec-n2p4m11S-McL}
\begin{split}
	&\ts b_3=1-2(b_1+b_2),\quad a_3=\frac{1}{2}-(a_1+a_2),\quad b_1=\frac{2}{5},\quad b_2=-\frac{1}{10},\\
	&\ts a_1 = \frac{1}{108}\left(14-\sqrt{19}\right)\approx 0.089269,\quad a_2 = \frac{1}{108}\left(20-7\sqrt{19}\right) \approx -0.097336
\end{split}
\end{equation}
which has error $\veps\approx 0.023685$ ($A<B$).

\myitemBest
\emph{Optimized ($m=11$, type S, $\nu=6$).} -- There are four constraints and hence two free parameters; according to the Gr\"{o}bner basis, we can choose $\{b_1,b_2\}$. The error is minimized for
\begin{subequations}
\begin{equation}\label{eq:dec-n2p4m11S-1}
\begin{split}
	&\ts b_3=1-2(b_1+b_2),\quad a_3=\frac{1}{2}-(a_1+a_2),\\
	& b_1=0.42652466131587616168,\quad b_2=-0.12039526945509726545,\\
	&\ts a_1 = 0.095848502741203681182,\quad a_2 = -0.078111158921637922695
\end{split}
\end{equation}
which gives $\veps\approx 0.018684$ ($B<A$). A nearby analytical solution with similar error ($\veps\approx 0.019991$) is
\begin{equation}\label{eq:dec-n2p4m11S-1A}
\begin{split}
	&\ts b_3=1-2(b_1+b_2),\quad a_3=\frac{1}{2}-(a_1+a_2),\quad b_1=\frac{3}{7},\quad b_2=-\frac{3}{25},\\
	&\ts a_1 = \frac{23 \left(25454-7 \sqrt{1125991}\right)}{4233384},\quad a_2 = \frac{91875-121 \sqrt{1125991}}{470376}.
\end{split}
\end{equation}
Another local minimum is located at
\begin{equation}\label{eq:dec-n2p4m11S-2}
\begin{split}
	&\ts b_3=1-2(b_1+b_2),\quad a_3=\frac{1}{2}-(a_1+a_2),\\
	& b_1=0.24759965401237406809,\quad b_2=-0.11679903600878927064,\\
	&\ts a_1 = 0.085676159176699987229,\quad a_2 = 0.49899422969605248140.
\end{split}
\end{equation}
It has error $\veps\approx 0.019074$ ($A<B$). See Fig.~\ref{fig:n2Errors}(c,d).
\end{subequations}

\myitem
\emph{Optimized ($m=13$, type SL, $\nu=3$).} -- There are two constraints and hence there is one free parameter; according to the Gr\"{o}bner basis, we can choose $w_2$. The error is minimized for
\begin{equation}\label{eq:dec-n2p4m13SL}\ts
	w_3=1-2(w_1+w_2),\quad
	w_1=\frac{1}{12} \left(4+2^{4/3}+2^{2/3}\right),\quad
	w_2 =-\frac{1}{6} \left(1+2^{1/3}\right)^2
\end{equation}
which gives $\veps\approx 0.28728$ ($A<B$).

\myitem
\emph{Optimized ($m=13$, type S, $\nu=7$).} -- There are four constraints and hence three free parameters; according to the Gr\"{o}bner basis, we can choose $\{a_2, b_2, a_3\}$. It is nontrivial to locate the global minimum. The best solution we found is
\begin{subequations}\label{eq:dec-n2p4m13S}
\begin{equation}\label{eq:dec-n2p4m13S-1}
\begin{split}
	&\ts a_4=1-2(a_1+a_2+a_3),\quad b_3=\frac{1}{2}-(b_1+b_2),\quad a_2 = 0.36781398298317937022,\\
	&\ts b_2 = -0.092981212295614937267,\quad a_3 = -0.068212103824011730130,\\
	&\ts a_1 = 0.074319284239746906187,\quad b_1 = 0.074319284239746906187
\end{split}
\end{equation}
which gives $\veps\approx 0.013886$ ($A<B$). A nearby analytical solution with similar error ($\veps\approx 0.014704$) is
\begin{equation}\label{eq:dec-n2p4m13S-1A}
\begin{split}
	&\ts a_4=1-2(a_1+a_2+a_3),\quad b_3=\frac{1}{2}-(b_1+b_2),\quad
	     a_2 = \frac{7}{19},\quad b_2 = -\frac{4}{43},\quad a_3 = -\frac{2}{29},\\
	&\ts a_1 = \frac{28509-4 \sqrt{14575449}-3 y}{142158},\quad b_1 = \frac{6487-y}{28380},
	\quad\text{with}\quad y=\sqrt{18920 \sqrt{14575449}-71143921}.
\end{split}
\end{equation}
\end{subequations}
\begin{figure}[t]
  \includegraphics[width=0.99\columnwidth]{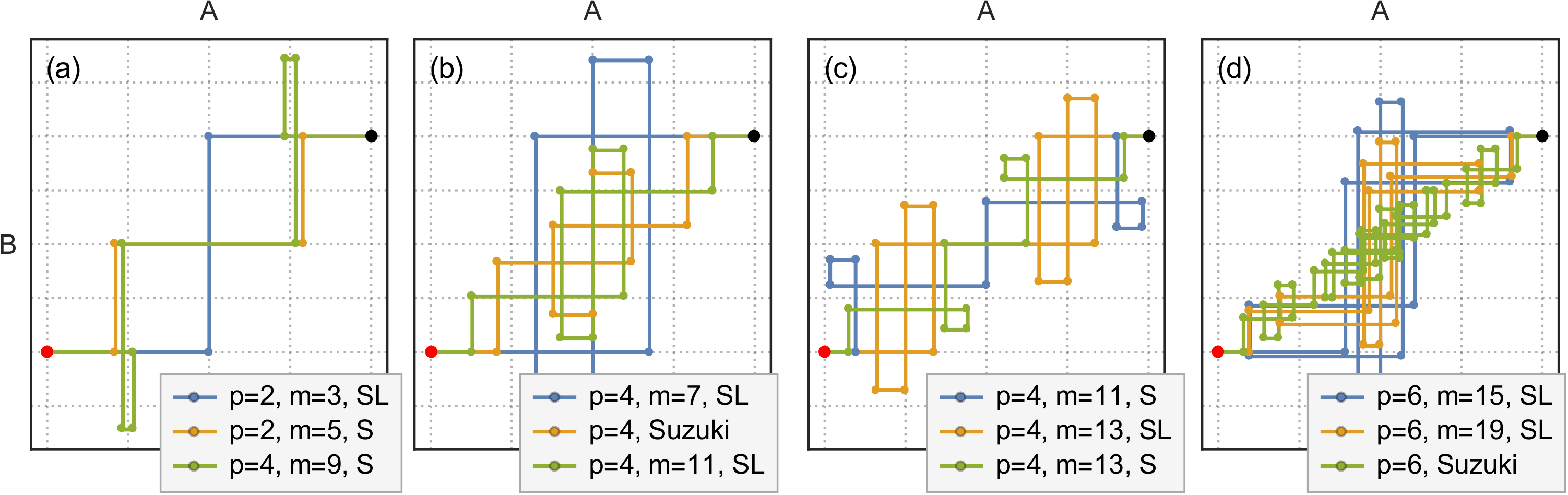}
  \caption{\label{fig:n2Decomps}Lie-Trotter-Suzuki decompositions for $n=2$ terms, i.e., $H=A+B$. The decompositions start at the red dots $(0,0)$ and then take steps $e^{a_i tA}$ and $e^{b_i tB}$ in the $A$ and $B$ ``directions'' until arriving at the black dot representing $(\sum_i a_i,\sum_i b_i)=(1,1)$.
  \textbf{(a)} The leapfrog decomposition (a.k.a.\ Verlet integrator), the recommended optimal second order type-S decomposition \eqref{eq:dec-n2p2Best} with $m=5$ factors, and the fourth order type-S decomposition \eqref{eq:dec-n2p4m9S-1}.
  \textbf{(b)} Fourth order type-SL decompositions: $U_{\tY,q=2}$ in Eq.~\eqref{eq:Yoshida} due to Forest, Ruth, and Yoshida with $m=7$ factors, $U_{\tZ,q=2}$ in Eq.~\eqref{eq:Suzuki} due to Suzuki with $m=11$, and the optimized decomposition \eqref{eq:dec-n2p4m11SL-1}, also with $m=11$ factors.
  \textbf{(c)} The recommended fourth order decomposition \eqref{eq:dec-n2p4m11S-1} with $m=11$, and the decompositions \eqref{eq:dec-n2p4m13SL} and \eqref{eq:dec-n2p4m13S} with $m=13$ factors.
  \textbf{(d)} Optimized sixth order decompositions \eqref{eq:dec-n2p6m19SL} and \eqref{eq:dec-n2p6m23SL} with $m=15$ and $m=19$ factors, respectively, and Suzuki's decomposition $U_{\tZ,q=3}$ in Eq.~\eqref{eq:Suzuki} with $m=51$.}
\end{figure}

\myitemSummary
\emph{Discussion.} -- The best decomposition found here is of type S with $m=13$ as specified in Eq.~\eqref{eq:dec-n2p4m13S}. It improves over the decomposition $U_{\tY,q=2}$ [Eq.~\eqref{eq:Yoshida}] of Forest, Ruth, and Yoshida \cite{Forest1990-43,Yoshida1990} by a factor $\sim 1/28$ and by a factor of $\sim 1/16$ over the widely applied decomposition $U_{\tZ,q=2}$ [Eq.~\eqref{eq:Suzuki}] due to Suzuki \cite{Suzuki1990}. Actually, the error of the decomposition \eqref{eq:dec-n2p4m13S} with $m=13$ factors does not improve too much over that of the type-S decomposition with $m=11$ factors [Eqs.~\eqref{eq:dec-n2p4m11S-1} and \eqref{eq:dec-n2p4m11S-1A}]. We have applied the latter in many tensor network simulations as in Refs.~\cite{Barthel2013-15,Lake2013-111,Cai2013-111,Barthel2016-94,Barthel2017_08unused,Binder2018-98} and recommend it generally for fourth order integration.
It is not surprising that the optimized type-SL decomposition \eqref{eq:dec-n2p4m13SL} with $m=13$ factors has a larger error than the SL decomposition \eqref{eq:dec-n2p4m11SL-1} with $m=11$ factors. The number of free parameters does not increase when going to $m=13$, but the increased number of factors is taken account of in the definition \eqref{eq:n2Error} of $\veps$ and results in a larger error value.
According to Table~\ref{tab:n2NoConstraints}, we can reach order $p=6$ with $m=15$ factors.

\subsection{Order \texorpdfstring{$p=6$}{p=6}}
\myitem
\emph{Yoshida ($m=15$, type SL, $\nu=4$).} -- There are four constraints and hence no free parameters. According to the Gr\"{o}bner basis, there are three real solutions which have already been determined numerically by Yoshida \cite{Yoshida1990}. Note that this is different from Yoshida's generic solution \eqref{eq:Yoshida} which has $m=19$ factors for $p=6$ and is discussed below. The best of the three solutions is
\begin{equation}\label{eq:dec-n2p6m15SL}
\begin{split}
	&w_4=1-2(w_1+w_2+w_3),\quad w_1=0.78451361047755726382,\\
	&w_2 =0.23557321335935813368,\quad w_3=-1.17767998417887100695
\end{split}
\end{equation}
which has error $\veps \approx 0.44573$ ($A<B$). The other two solutions have much larger errors $\veps \approx 5.7167$ and $\veps \approx 5.8160$, respectively (both for order $A<B$).

\myitem
\emph{Yoshida, Kahan \& Li ($m=19$, type SL).} -- For Yoshida's decomposition $U_{\tY,q=3}$ with $q=3$ in Eq.~\eqref{eq:Yoshida}, the error is $\veps \approx 26.18692$ ($A<B$). In Ref.~\cite{Kahan1997-66}, Kahan and Li state two similar solutions. The better of the two is
\begin{align}\nonumber
	&w_5=1-2(w_1+w_2+w_3+w_4),\,\, w_1=0.3910302033086847882,\,\, w_2 =0.3340372896111360175,\\
	\label{eq:dec-n2p6m19SL-KahanLi}
	&w_3=-0.70622728118756134346,\quad w_4=0.081877549648059445768
\end{align}
with error $\veps\approx 0.22167$ ($A<B$).

\myitemBest
\emph{Optimized ($m=19$, type SL, $\nu=5$).} -- There are four constraints and hence there is one free parameter; according to the Gr\"{o}bner basis, we can choose $w_2$. The error is minimized for
\begin{equation}\label{eq:dec-n2p6m19SL}
\begin{split}
	&w_5=1-2(w_1+w_2+w_3+w_4),\quad w_1=0.18793069262651671457,\quad w_2 =0.5553,\\
	&w_3=0.12837035888423653774,\quad w_4=-0.84315275357471264676
\end{split}
\end{equation}
which gives $\veps\approx 0.17255$ ($B<A$). See Fig.~\ref{fig:n2Errors}(e). Albeit optimizing a different error measure, McLachlan gave almost the same decomposition in Ref.~\cite{McLachlan1995}.
\begin{figure}[t]
  \includegraphics[height=0.23\columnwidth]{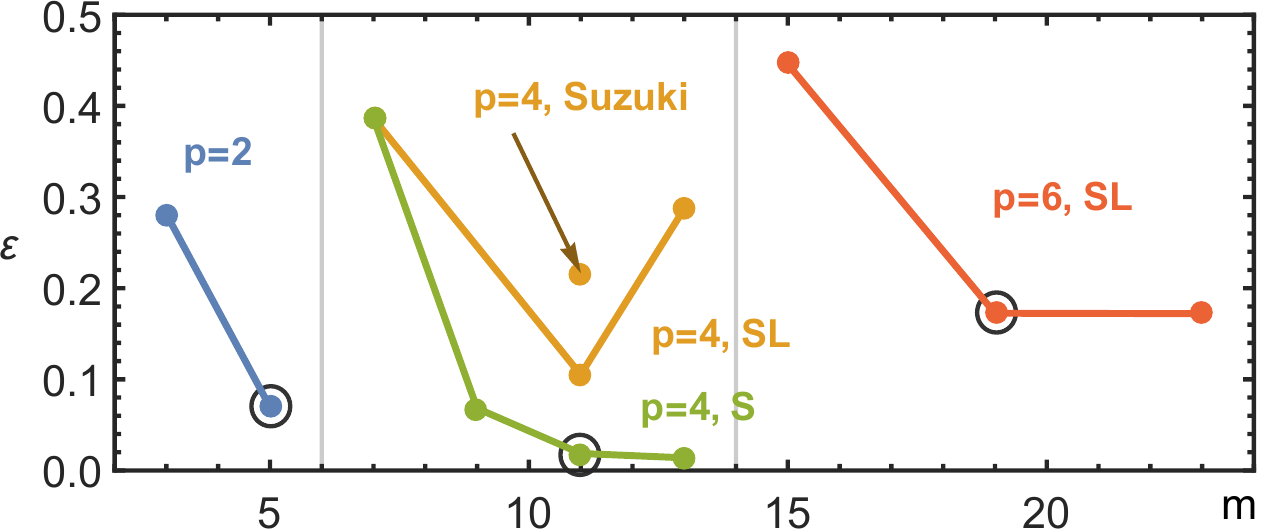}
  \caption{\label{fig:n2_errors}Error values $\veps$ for the optimized $n=2$ decompositions specified in Sec.~\ref{sec:n2opt}, and Suzuki's widely used fourth order decomposition \eqref{eq:Suzuki}. Yoshida's and Suzuki's unoptimized sixth order integrators are not displayed because of their large errors. Recommended decompositions for each order are indicated by circles. As discussed in Sec.~\ref{sec:n2ErrorMeasure}, error values of decompositions with different order $p$ can not be compared directly.}
\end{figure}

\myitem
\emph{Optimized ($m=23$, type SL, $\nu=6$).} -- There are four constraints and hence there are two free parameters; according to the Gr\"{o}bner basis, we can choose $\{w_4,w_5\}$. The best solution we found has
\begin{equation}\label{eq:dec-n2p6m23SL}
\begin{split}
	&w_6=1-2(w_1+w_2+w_3+w_4+w_5),\quad w_1=0.11246183971085248218,\quad\\
	&w_2=0.21955991439348897340,\quad w_3=0.47486253551971306793,\quad\\
	&w_4=-0.74,\quad w_5=0.018
\end{split}
\end{equation}
which gives $\veps\approx 0.17204$ ($A<B$). We identified two further good local minima with $\veps\approx 0.19599$ ($B<A$) and $\veps\approx 0.26551$ ($B<A$), respectively. See Fig.~\ref{fig:n2Errors}(f).

\emph{Suzuki ($m=51$, type SL).} -- For the decomposition $U_{\tZ,q=3}$ in Eq.~\eqref{eq:Suzuki}, the error is $\veps\approx 0.84749 $ ($B<A$). The order $A<B$ for the Hall basis would give a considerably larger error of $\approx 16.992$.

\myitemSummary
\emph{Discussion.} -- The best decomposition found here is of type SL with $m=23$ as specified in Eq.~\eqref{eq:dec-n2p6m23SL}. It improves over Yoshida's generic decomposition $U_{\tY,q=3}$ [Eq.~\eqref{eq:Yoshida}] by a factor $\sim 1/152$, by a factor of $\sim 1/5$ over Suzuki's decomposition $U_{\tZ,q=3}$ [Eq.~\eqref{eq:Suzuki}], and by a factor of $\sim 2/5$ over the solution \eqref{eq:dec-n2p6m15SL} that Yoshida found numerically \cite{Yoshida1990}. Actually, the error of the decomposition \eqref{eq:dec-n2p6m23SL} with $m=23$ factors is almost identical to that of the decomposition \eqref{eq:dec-n2p6m19SL} with $m=19$ factors. We hence recommend using the latter for sixth order integration.
According to Table~\ref{tab:n2NoConstraints}, true type-S decompositions require $m=19$ factors to reach order $p=6$. We have checked explicitly that type-S solutions with $m=15$ simply reproduce the corresponding type-SL decomposition \eqref{eq:dec-n2p6m15SL}. With $m=19$, type-S decompositions should just reproduce the type-SL decomposition \eqref{eq:dec-n2p6m19SL} with one free parameter. With $m=23$, type-S decompositions have two free parameters like the corresponding type-SL decompositions \eqref{eq:dec-n2p6m23SL}. According to Table~\ref{tab:n2NoConstraints}, there exist type-SL decompositions of order $p=8$ for $m\geq 31$ factors.

\section{Lie-Trotter-Suzuki decompositions for \texorpdfstring{$n=3$}{n=3} terms}\label{sec:n3}
For $n=3$, the generator $H=A+B+C$ consists of three terms, where exponentials $e^{tA}$, $e^{tB}$, $e^{tC}$ can be computed easily. Examples for corresponding partitionings of lattice model interaction graphs are given in Fig.~\ref{fig:partition}(b,d,e,f). Much of the treatment for two terms in Sec.~\ref{sec:n2} carries over, but there are also some differences. In particular, we have two further relevant types of symmetric decompositions.

\subsection{Considered types of decompositions}\label{sec:n3Decomps}
We consider the following types of Lie-Trotter-Suzuki decompositions for $e^{tH}$, where $m$ denotes the total number of operator exponentials:
\begin{itemize}
  \item
  \textbf{Type N} with $\nu=m$ parameters. This is the most generic type of decompositions with the building block $ABCB$
  \begin{equation}\label{eq:n3TypeN}
  	U_{\tN,m}:=
  	e^{a_1 tA}\,e^{b_1 tB}\,e^{c_1 tC}\,e^{b_2 tB}\,e^{a_2 tA}\,e^{b_3 tB}\,e^{c_2 tC}\,e^{b_4 tB}\dotsb
  \end{equation}
  with $a_i,b_i,c_i\in\mathbb{R}$.
  \item
  \textbf{Type S} with $\nu=\ceil(m/2)=(m+1)/2$ parameters and even $(m-1)/2$. This is a symmetric decomposition with building block $ABCB$:
  \begin{equation}\label{eq:n3TypeSabcb}
  	U_{\tS,m}:=
  	e^{a_1 tA}\,e^{b_1 tB}\,e^{c_1 tC}\,e^{b_2 tB}\,e^{a_2 tA}\,e^{b_3 tB}\,e^{c_2 tC}\,e^{b_4 tB}\,\dotsb e^{b_2 tB}\,e^{c_1 tC}\,\,e^{b_1 tB}\,e^{a_1 tA}
  \end{equation}
  with $a_i,b_i,c_i\in\mathbb{R}$.
  \item
  \textbf{Type S-abc} with $\nu=\ceil(m/2)$ parameters and even $(m+1)/3$. This is a symmetric decomposition with building block $ABC$:
  \begin{equation}\label{eq:n3TypeSabc}
  	U_{\tS',m}:=
  	e^{a_1 tA}\,e^{b_1 tB}\,e^{c_1 tC}\,e^{a_2 tA}\,e^{b_2 tB}\,e^{c_2 tC}\dotsb e^{c_2 tC}\,\,e^{b_2 tB}\,e^{a_2 tA}\,e^{c_1 tC}\,\,e^{b_1 tB}\,e^{a_1 tA}
  \end{equation}
  with $a_i,b_i,c_i\in\mathbb{R}$.
  \item
  \textbf{Type SE} with $\nu=(m-1)/4$ parameters and even $(m-1)/2$. This type of symmetric decomposition is a product of alternating two types of ``Euler'' terms \footnote{The name alludes to the similarity to the Euler integration method for differential equations. This first order decomposition is the famous Lie-Trotter product formula \cite{Trotter1959}.}
  \begin{equation}\label{eq:Euler}
  	U_\tEp(\tau):= e^{\tau A}\,e^{\tau B}\,e^{\tau C}\quad\text{and}\quad
  	U_\tEm(\tau):= e^{\tau C}\,e^{\tau B}\,e^{\tau A}
  \end{equation}
   such that 
  \begin{align}\nonumber
  	U_{\tSE,m}&:=U_\tEp(u_1t)\,U_\tEm(v_1t)\,U_\tEp(u_2t)\,U_\tEm(v_2t)\,U_\tEp(u_3t)\,U_\tEm(v_3t)\dotsb  U_\tEp(v_1t)\,U_\tEm(u_1t)\\
  	&\phantom{:}=e^{u_1 tA}\,e^{u_1 tB}\,e^{(u_1+v_1) tC}\,e^{v_1 tB}\,e^{(v_1+u_2) tA}\,e^{u_2 tB}\,e^{(u_2+v_3) tC}
  	         \dotsb e^{u_1 tB}\,e^{u_1 tA}
  	\label{eq:n3TypeSE}
  \end{align}
  with $u_i,v_i\in\mathbb{R}$. Note that, as shown in the last line, exponentials of $C$ or $A$ for subsequent Euler terms $U_\tEp(ut)U_\tEm(vt)$ and $U_\tEm(vt)U_\tEp(ut)$, respectively, can be contracted into one such that the total number of required operator exponentials is indeed $m$. For $n=2$ terms, we did not discuss this type of decompositions because it is in that case simply equivalent to type S in Eq.~\eqref{eq:n2TypeS} \cite{McLachlan1995}.
 \item
  \textbf{Type SL} with $\nu=\ceil((m-1)/8)$ parameters and integer $(m-1)/4$. This type of symmetric decomposition is a product of leapfrog terms
  \begin{equation}\label{eq:leapfrog3}
  	U_{\tL}(\tau):= e^{\frac{1}{2}\tau A}\,e^{\frac{1}{2}\tau B}\,e^{\tau C}\,e^{\frac{1}{2}\tau B}\,e^{\frac{1}{2}\tau A}
  \end{equation}
   such that 
  \begin{align}\nonumber
  	U_{\tSL,m}&:= U_\tL(w_1t)\,U_\tL(w_2t)\,U_\tL(w_3t)\dotsb U_\tL(w_{\nu}t) \dotsb U_\tL(w_1t)\\
  	&\phantom{:}=e^{\frac{1}{2}w_1 tA}\,e^{\frac{1}{2}w_1 tB}\,e^{w_1 tC}\,e^{\frac{1}{2}w_1 tB}\,e^{\frac{1}{2}(w_1+w_2) tA}\,e^{\frac{1}{2}w_2 tB}\,e^{w_2 tC}\dotsb e^{\frac{1}{2}w_1 tA} \label{eq:n3TypeSL}
  \end{align}
  with $w_i\in\mathbb{R}$. Note that, as shown in the last line, exponentials of $A$ for subsequent leapfrog terms can be contracted into one such that the total number of required operator exponentials is indeed $m$.
\end{itemize}

Depending on the number $m$ of factors in a decomposition $U$ of type \eqref{eq:n3TypeN}-\eqref{eq:n3TypeSL}, the parameters can be chosen such that $U$ coincides with the exact $e^{tH}$ up to order $p$ in the sense of Eq.~\eqref{eq:constraint}.
If free parameters remain, we can use these to minimize the leading error term of the decomposition.

Of course, for the same number of factors $m$, some types of decompositions are subclasses of others:
\begin{equation}
	\tSL\subseteq\tSE\subseteq\tS\subseteq\tN \quad\text{and}\quad
	\text{S-abc}\subseteq\tS\subseteq\tN.
\end{equation}

\subsection{Numbers of parameters and constraints, symmetries}\label{sec:n3NoParam}
As in Eqs.~\eqref{eq:BCHiter}, we can use the BCH formula recursively, to compute $\log U$ in terms of nested commutators of $A$, $B$, and $C$. The number of constraints, imposed by requiring $U$ to coincide with $e^{tH}$ up to order $t^{p}$ [Eq.~\eqref{eq:constraint2}], can be determined from the number of terms in the expansion. For the large Hilbert spaces relevant in many-body physics and low expansion orders considered for our optimized decompositions, it is sufficient to treat the Lie algebra generated by $\{A,B,C\}$ as free. We can hence work with Hall bases \cite{Hall1950-1,Serre1992,Reutenauer1993} as discussed in Sec.~\ref{sec:n2HallBases}. The number of Hall basis elements of degree $k$ is given by the necklace polynomial \eqref{eq:HallBasisDim}. Tables~\ref{tab:HallBasis} and \ref{tab:HallBasisDim} list Hall basis elements and their numbers $d_k$.

Assuming that all terms that can occur do occur in the expansion of $Z=\log U$ for an order-$p$ Lie-Trotter-Suzuki decomposition $U$, we read off constraint polynomials as coefficients of the relevant Hall basis elements. To check and determine how many free parameters we actually have, we can employ the Gr\"{o}bner basis of the constraint polynomials as discussed in Sec.~\ref{sec:Groebner}.

\textbf{Type N.} -- For the decompositions $U_{\tN,m}$ in Eq.~\eqref{eq:n3TypeN}, we have $\nu=m$ parameters $\{a_i,b_i,c_i\}$. As the decomposition has no further symmetry or structure, all Hall basis elements should occur in the expansion of $Z$ and the number of constraints to achieve approximation order $p$ is $\sum_{k=1}^p d_k$. For example, at first order, we have the three constraints $\sum_i a_i=\sum_i b_i=\sum_i c_i = 1$ to achieve $Z=tH+\mc{O}(t^2)$.

\textbf{Type S}. -- For the decompositions $U_{\tS,m}$ in Eq.~\eqref{eq:n3TypeSabcb}, we have $\nu=(m+1)/2$ parameters and even $(m-1)/2$. Due to the symmetry in the factors, the decompositions obey the time reversal symmetry $U_{\tS,m}(t)U_{\tS,m}(-t)=\id$. According to Lemma~\ref{lemma:n2ReversalSym}, it follows that $Z(t)$ only contains terms of odd order in $t$ such that no Hall basis elements of even degree occur \cite{Yoshida1990}. For a symmetric decomposition that has no further structure, all Hall basis elements of odd degree should occur in the expansion of $Z$ and the number of constraints to achieve approximation order $p$ is $\sum_{q,2q-1\leq p} d_{2q-1}$. For example, at first order, we have the three constraints that the coefficients $a_i$ in the $A$ factors, $b_i$ in the $B$ factors and, $c_i$ in the $C$ factor of Eq.~\eqref{eq:n3TypeSabcb} sum to one to achieve $Z=tH+\mc{O}(t^2)$.

\textbf{Type S-abc}. -- For the decompositions $U_{\tS',m}$ in Eq.~\eqref{eq:n3TypeSabc}, we have $\nu=\ceil(m/2)$ parameters and even $(m+1)/3$. As these decompositions are symmetric, only terms of odd degree occur in the expansion of $Z=\log U_{\tS',m}$. As for type S, the number of constraints to achieve approximation order $p$ is $\sum_{q,2q-1\leq p} d_{2q-1}$.
\begin{table}[t]
	\setlength{\tabcolsep}{1.5ex}
	\begin{tabular}{c | l}
	 Degree & Hall basis elements\\
	 \hline
	 1	& $A_1$\\
	 3	& $A_3$,\, $[A_1,A_2]$\\
	 5	& $A_5$,\, $[A_1,A_4]$,\, $[A_2,A_3]$,\, $[A_1,[A_1,A_3]]$,\, $[A_2,[A_1,A_2]]$,\, $[A_1,[A_1,[A_1,A_2]]]$\\
	 7	& $A_7$,\, $[A_1,A_6]$,\, $[A_2,A_5]$,\, $[A_3,A_4]$,\, $[A_1,[A_1,A_5]]$,\, $[A_2,[A_1,A_4]]$,\,\\
	 	& $[A_2,[A_2,A_3]]$,\, $[A_3,[A_1,A_3]]$,\, $[A_4,[A_1,A_2]]$,\, $[A_1,[A_1,[A_1,A_4]]]$,\,\\
	 	& $[A_2,[A_1,[A_1,A_3]]]$,\, $[A_2,[A_2,[A_1,A_2]]]$,\, $[A_3,[A_1,[A_1,A_2]]]$,\, $[[A_1,A_2],[A_1,A_3]]$,\,\\
	 	& $[A_1,[A_1,[A_1,[A_1,A_3]]]]$,\, $[A_2,[A_1,[A_1,[A_1,A_2]]]]$,\, $[[A_1,A_2],[A_1,[A_1,A_2]]]$,\,\\
	 	& $[A_1,[A_1,[A_1,[A_1,[A_1,A_2]]]]]$
	\end{tabular}
	\caption{\label{tab:HallBasisE}Hall basis elements of odd degree for the free Lie algebra generated by operators $\{A_1,A_2,A_3,A_4,\dotsc\}$, where $A_k$ has degree $k$.}
\end{table}
\begin{table}[t]
	\setlength{\tabcolsep}{1.5ex}
	\begin{tabular}{c | c c c c c c c c c c c c}
	 Degree $k$  & 1 & 3 & 5 & 7 & 9 & 11 & 13\\
	 \hline
	 $d_{\tE,k}$        & 1 & 2 & 6 & 18 & 56 & 186 & 630
	\end{tabular}
	\caption{\label{tab:HallBasisEDim}Numbers $d_{\tE,k}$ of Hall bases elements with odd degree $k$ for the free Lie algebra generated by operators $\{A_1,A_2,A_3,A_4,\dotsc\}$, where $A_k$ has degree $k$.}
\end{table}
\begin{table}[t]
	\setlength{\tabcolsep}{1.5ex}
	Number $N^\tH_p$ of Hall basis elements with degree $\leq p$ in an expansion of $\log U$ with $n=3$:\\[0.5em]
	\begin{tabular}{c || c | c | c | c | c | c | c | c | c | c}
	 order $p$   & 1 & 2 & 3 & 4 & 5 & 6 & 7 & 8 & 9 & 10\\
	 \hline
 	 Type N     & 3 & 6 & 14 & 32 & 80 & 196 & 508 & 1318 & 3502 & 9382\\
	 Type S     & --& 3 & -- & 11 & -- & 59  & --  & 371  &  --  & 2555\\
	 Type S-abc & --& 3 & -- & 11 & -- & 59  & --  & 371  &  --  & 2555\\
	 Type SE    & --& 1 & -- & 3  & -- & 9   & --  & 27   &  --  & 83\\
	 Type SL    & --& 1 & -- & 2  & -- & 4   & --  &  8   &  --  & 16
	\end{tabular}\\[1em]
	Minimum number $m_p^\tH$ of factors needed to allow for  $\log U= tH + \mc{O}(t^{p+1})$ with $n=3$:\\[0.5em]
	\begin{tabular}{c || c | c | c | c | c | c | c | c | c | c}
	 order $p$      & 1 & 2 & 3 & 4 & 5 & 6 & 7 & 8 & 9 & 10\\
	 \hline
 	 Type N     & 3 & 6 & 14 & 32 & 80 & 196 & 508 & 1318 & 3502 & 9382\\
	 Type S     & --& 5 & -- & 21 & -- & 117 & --  & 741  &  --  & 5109\\
	 Type S-abc & --& 5 & -- & 23 & -- & 119 & --  & 743  &  --  & 5111\\
	 Type SE    & --& 5 & -- & 13 & -- & 37  & --  & 109  &  --  & 333\\
	 Type SL    & --& 5 & -- & 13 & -- & 29  & --  & 61   &  --  & 125
	\end{tabular}
	\caption{\label{tab:n3NoConstraints}For the decompositions $U$ defined in Sec.~\ref{sec:n3Decomps} with $n=3$ non-commuting terms in $H$, the first table gives the number $N^\tH_p$ of Hall basis elements with degree $\leq p$ in an expansion of $\log U$. Assuming that the resulting $N^\tH_p$ constraints to obtain an order-$p$ decomposition of $e^{tH}=e^{t(A+B+C)}$ are independent, the second table gives the corresponding minimum number $m_p^\tH$ of factors needed to for order-$p$ decompositions.}
\end{table}

\textbf{Type SE.} -- For the decompositions $U_{\tSE,m}$ in Eq.~\eqref{eq:n3TypeSE}, we have $\nu=(m-1)/4$ parameters and even $(m-1)/2$. As these decompositions are symmetric, only terms of odd degree occur in the expansion of $Z=\log U_{\tSE,m}$. The decomposition is a product of Euler terms $U_\tEp(\tau)$ and $U_\tEm(\tau)$ in Eq.~\eqref{eq:Euler} which are non-symmetric first order decompositions of $e^{\tau H}$ \cite{Trotter1959}. To determine the number of constraints, we can use the following expansions.
\begin{lemma}[\lemmaHead{Euler term expansions}]\label{lemma:Euler}
  The expansions of the forward and backward Euler terms $U_{\tE+}(\tau)=e^{\tau A_1}\,e^{\tau A_2}\dotsb e^{\tau A_n}$ and $U_{\tE-}(\tau)=e^{\tau A_n}\,e^{\tau A_{n-1}}\dotsb e^{\tau A_1}$ coincide up to sign factors. In particular,
  \begin{equation}\label{eq:EulerTerms} \ts
  	\log U_{\tE\pm}(\tau)=\tau Z^{(1)}_\tE\mp \tau^2 Z^{(2)}_\tE+ \tau^3 Z^{(3)}_\tE\mp \tau^4 Z^{(4)}_\tE+\dots
  	\quad\text{with}\quad Z^{(1)}_\tE=\sum_{i=1}^n A_i.
  \end{equation}
\end{lemma}
For the proof, note that $U_{\tE+}(\tau)U_{\tE-}(-\tau)=\id$ and let us define $\log U_{\tE\pm}(\tau)=:\tau Z^{(1)}_\tE + \tau^2 Z^{(2)}_\pm+ \tau^3 Z^{(3)}_\pm+ \tau^4 Z^{(4)}_\pm+\dots$ where $Z^{(1)}_\tE=\sum_{i=1}^n A_i$. Applying the BCH formula \eqref{eq:BCH1}, we find $\log[U_{\tE+}(\tau)U_{\tE-}(-\tau)]=\tau ^2 (Z^{(2)}_++Z^{(2)}_-)+\mc{O}(\tau^3)$ such that $Z^{(2)}_+=-Z^{(2)}_-=:Z^{(2)}_{\tE +}$ because of $U_{\tE+}(\tau)U_{\tE-}(-\tau)=\id$. With this information, we can reconsider the BCH formula and find $\log[U_{\tE+}(\tau)U_{\tE-}(-\tau)]=\tau^3 (Z^{(3)}_+-Z^{(3)}_-)+\mc{O}(\tau^4)$, showing that $Z^{(3)}_+=Z^{(3)}_-=:Z^{(3)}_{\tE +}$. Continuing in this way, Eq.~\eqref{eq:EulerTerms} is established.

$Z^{(k)}_\tE$ is a Lie polynomial containing only nested commutators of degree $k$. To determine the number of constraints for $U_{\tSE,m}$ to be an order-$p$ decomposition of $e^{t H}$, we can consider the Lie algebra generated by $\{Z^{(1)}_\tE,Z^{(2)}_\tE,Z^{(3)}_\tE,\dots\}$ as free. The number of constraints due to Eq.~\eqref{eq:constraint2} is then given by the number $\sum_{q,2q-1\leq p} d_{\tE,2q-1}$ of Hall basis elements of odd degree $2q-1\leq p$. Tables~\ref{tab:HallBasisE} and \ref{tab:HallBasisEDim} list the Hall basis elements and their numbers $d_{\tE,k}$.  At first order, we only have the constraint that the coefficients $u_i$ and $v_i$ of all Euler factors in Eq.~\eqref{eq:n3TypeSE} sum to one to achieve $Z=tH+\mc{O}(t^2)$.

\textbf{Type SL.} -- For the decompositions $U_{\tSL,m}$ in Eq.~\eqref{eq:n3TypeSL}, we have $\nu=\ceil((m-1)/8)$ parameters and integer $(m-1)/4$. As these decompositions are symmetric, only terms of odd degree occur in the expansion of $Z=\log U_{\tSL,m}$. The decomposition is a product of leapfrog terms $U_\tL(\tau)$ in Eq.~\eqref{eq:leapfrog3} which are symmetric second order decompositions of $e^{\tau H}$ and hence can be expanded in the form $\log U_\tL= \tau Z^{(1)}_\tL + \tau^3 Z^{(3)}_\tL + \tau^5 Z^{(5)}_\tL+ \dots$ with $Z^{(1)}_\tL=H$ [Eq.~\eqref{eq:leapfrogTerms}].
$Z^{(k)}_\tL$ is a Lie polynomial containing only nested commutators of degree $k$. To determine the number of constraints for $U_{\tSL,m}$ to be an order-$p$ decomposition of $e^{t H}$, we can consider the Lie algebra generated by $\{Z^{(1)}_\tL,Z^{(3)}_\tL,Z^{(5)}_\tL,\dots\}$ as free. The number of constraints due to Eq.~\eqref{eq:constraint2} is then given by the number $\sum_{q,2q-1\leq p} d_{\tL,2q-1}$ of Hall basis elements of odd degree $2q-1\leq p$. Tables~\ref{tab:HallBasisL} and \ref{tab:HallBasisLDim} list the Hall basis elements and their numbers $d_{\tL,k}$.  At first order, we only have the constraint that the coefficients $w_i$ of all leapfrog factors in Eq.~\eqref{eq:n3TypeSL} sum to one to achieve $Z=tH+\mc{O}(t^2)$.

To summarize this section, we give the total numbers of constraints by order in Table~\ref{tab:n3NoConstraints}. The table also states $m_p^\tH$, the minimum number of factors needed in the different decompositions to obtain an order-$p$ decomposition [cf.\ Eq.~\eqref{eq:constraint}], under the assumption that all constraint polynomials are independent.

\section{Optimized decompositions for \texorpdfstring{$n=3$}{n=3} terms}\label{sec:n3opt}
As discussed in Sec.~\ref{sec:n2ErrorMeasure} we can quantify the accuracy of an order-$p$ decomposition by the leading error term which is of order $t^{p+1}$. The relevant error measure is the operator norm distance which we bound using the triangle inequality. This leads to Eq.~\eqref{eq:normDiff} and the 1-norm $\sum_i |c_i|$ of the Hall basis expansion coefficients $|c_i|$ as an error measure. Furthermore, we are free to impose any order for the generators $A,B,C$ in the construction of the Hall basis and the $c_i$ depend on that choice. Let the sets of coefficient polynomials for the six possible orders be denoted by $\{c^{ABC}_i\}$, $\{c^{BCA}_i\}$ etc. We will then use the error measure
\begin{equation}\label{eq:n3Error}\ts
	\veps := \left(\frac{m}{p}\right)^p\min\left( \sum_i |c^{ABC}_i|,\, \sum_i |c^{BCA}_i|,\, \sum_i |c^{CAB}_i|,\, \sum_i |c^{BAC}_i|,\, \sum_i |c^{ACB}_i|,\, \sum_i |c^{CBA}_i|\right)
\end{equation}
to quantify the magnitude of the deviation of $U$ from $e^{tH}$, where $m$ denotes again the number of factors in the decomposition $U$. Like the coefficient polynomials $c_i$, the error $\veps$ is a function of the parameters in the decomposition.

Here, we optimize the decompositions of types N, S, S-abc, SE, and SL defined in Sec.~\ref{sec:n3Decomps} with respect to their parameters to minimize the error measure \eqref{eq:n3Error}. We compare the results to the unoptimized type-SL decompositions \eqref{eq:Yoshida} and \eqref{eq:Suzuki} of Yoshida \cite{Yoshida1990} and Suzuki \cite{Suzuki1990} which generalize without modification to arbitrary numbers $n$ of terms in $H$. Similarly, we compare to type-SL decompositions of McLachlan \cite{McLachlan1995}, adapted to the $n=3$ case, and those of Kahan and Li \cite{Kahan1997-66}.

In the discussion below, $\nu$ denotes the number of parameters as stated in Sec.~\ref{sec:n3Decomps} and the number of constraints for the different decomposition types and approximation orders $p$ are given in Table~\ref{tab:n3NoConstraints}. We check the applicability of the corresponding counting argument using Gr\"{o}bner bases for the constraint polynomials as discussed in Sec.~\ref{sec:Groebner}. For each order $p$, the best decomposition found (smallest $\veps$) is indicated by a star. When the different possible orders $A<B<C$, $B<C<A$ etc., chosen for the construction of the Hall basis, are not equivalent, we specify in brackets the one that yields the minimum for $\veps$.

\subsection{Order \texorpdfstring{$p=1$}{p=1}}
\myitemBest
\emph{Euler ($m=3$, type N, $\nu=1$).} -- All three parameters are fixed to $a_1=b_1=c_1=1$ by the constraint that the first order term in $\log U$ is $tH=t(A+B+C)$. The error is $\veps = \left(\frac{3}{1}\right)^1\frac{3}{2} = \frac{9}{2}$.

\subsection{Order \texorpdfstring{$p=2$}{p=2}}
\begin{figure}[p]
  \begin{tabular}{l@{\hspace{7ex}}l}
   {\small{(a)}}\quad \underline{$p=2$, $m=9$, type S}
  &{\small{(b)}}\quad \underline{$p=4$, $m=17$, type SE}\\[-0em]
                \imagetop{\includegraphics[height=0.25\columnwidth]{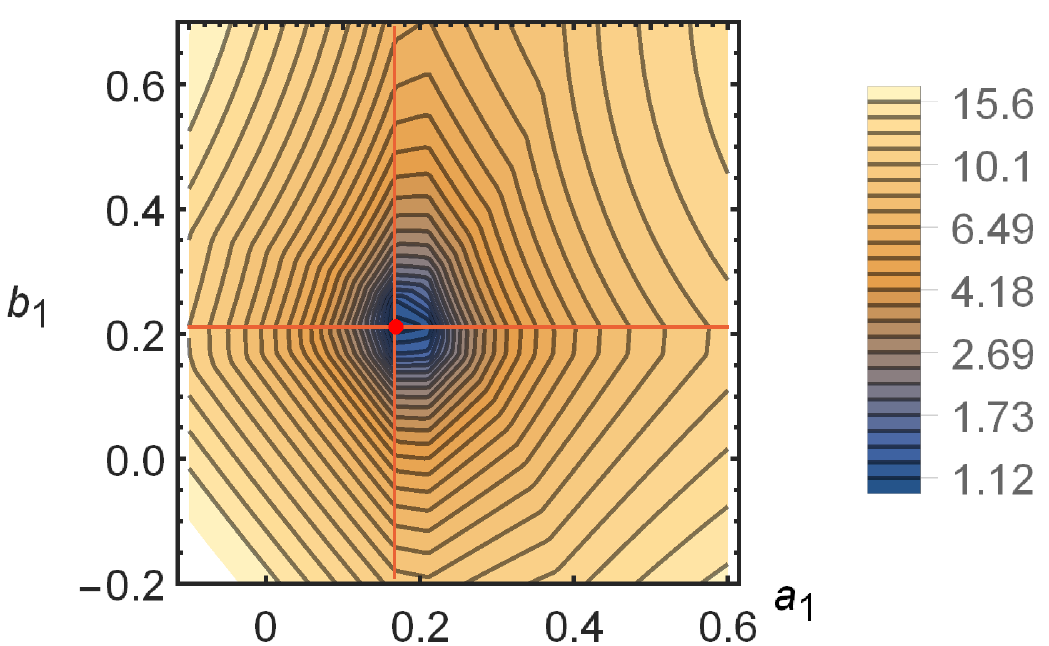}}\vspace{1.2em}
  &             \imagetop{\includegraphics[height=0.23\columnwidth]{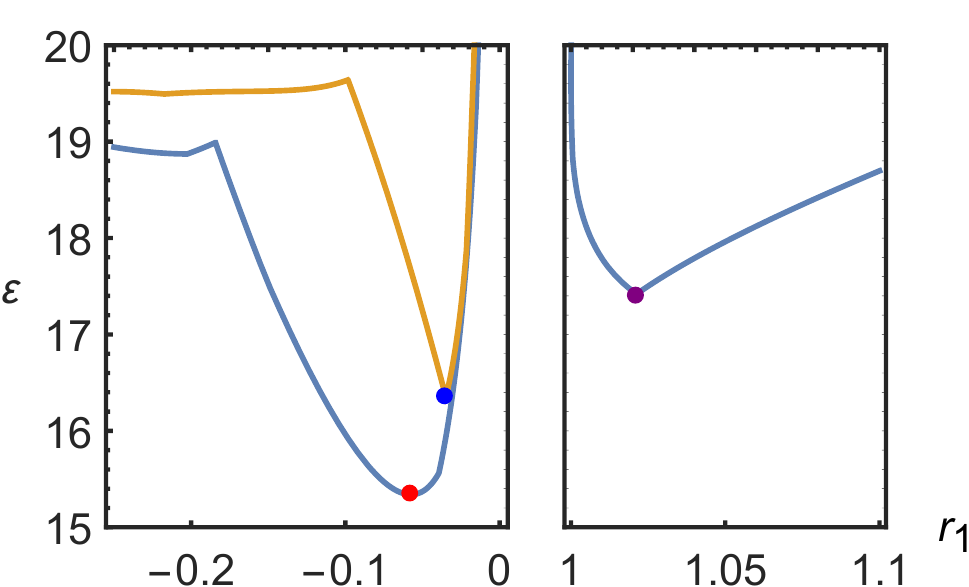}}\\[0.2em]
   {\small{(c)}}\quad \underline{$p=4$, $m=21$, type SE}
  &{\small{(d)}}\quad \underline{$p=6$, $m=37$, type SL}\\[-0em]
                \imagetop{\includegraphics[height=0.25\columnwidth]{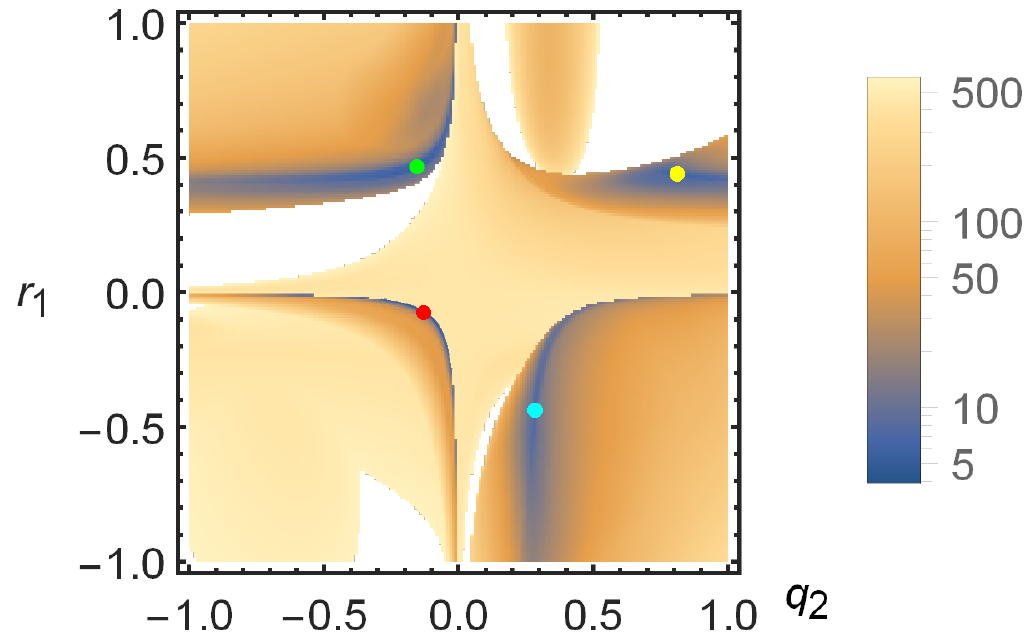}}\vspace{1.2em}
  &             \imagetop{\includegraphics[height=0.23\columnwidth]{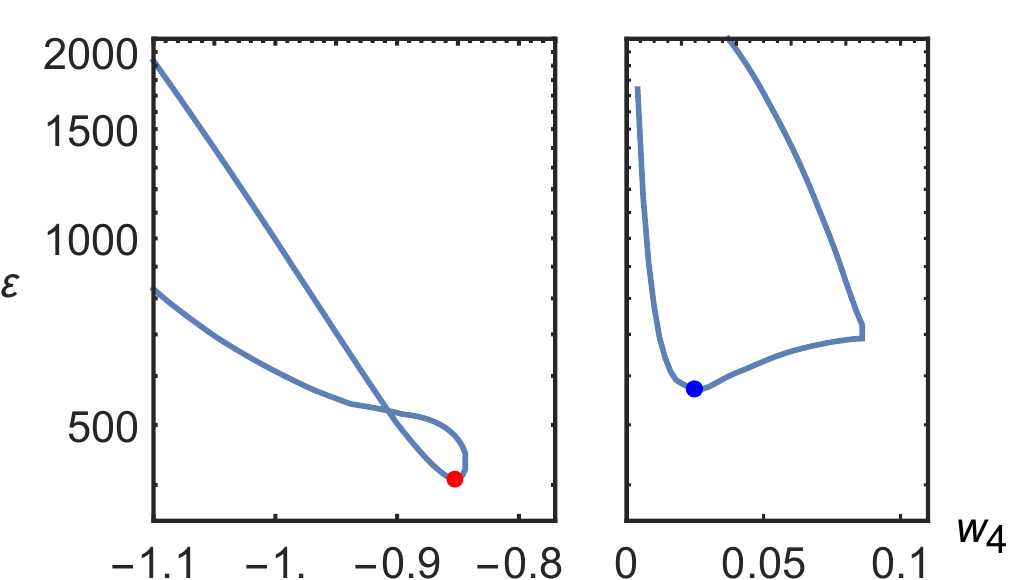}}\\[0.2em]
   {\small{(e)}}\quad \underline{$p=4$, $m=25$, type SE}
  &{\small{(f)}}\quad \underline{$p=4$, $m=25$, type SE}\\[-0em]
                \imagetop{\includegraphics[height=0.25\columnwidth]{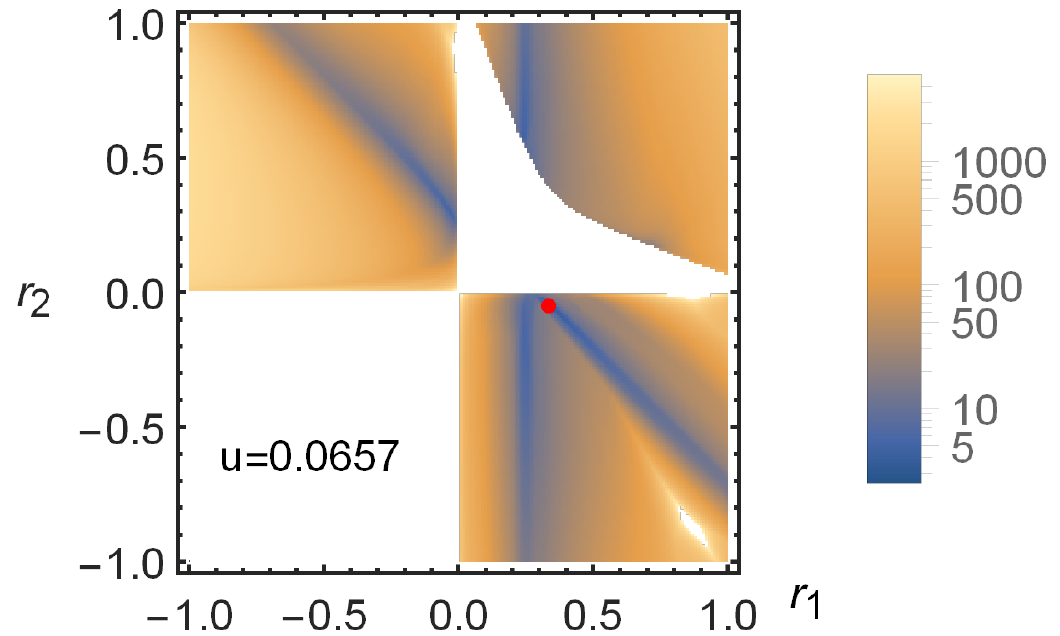}}\vspace{1.2em}
  &             \imagetop{\includegraphics[height=0.25\columnwidth]{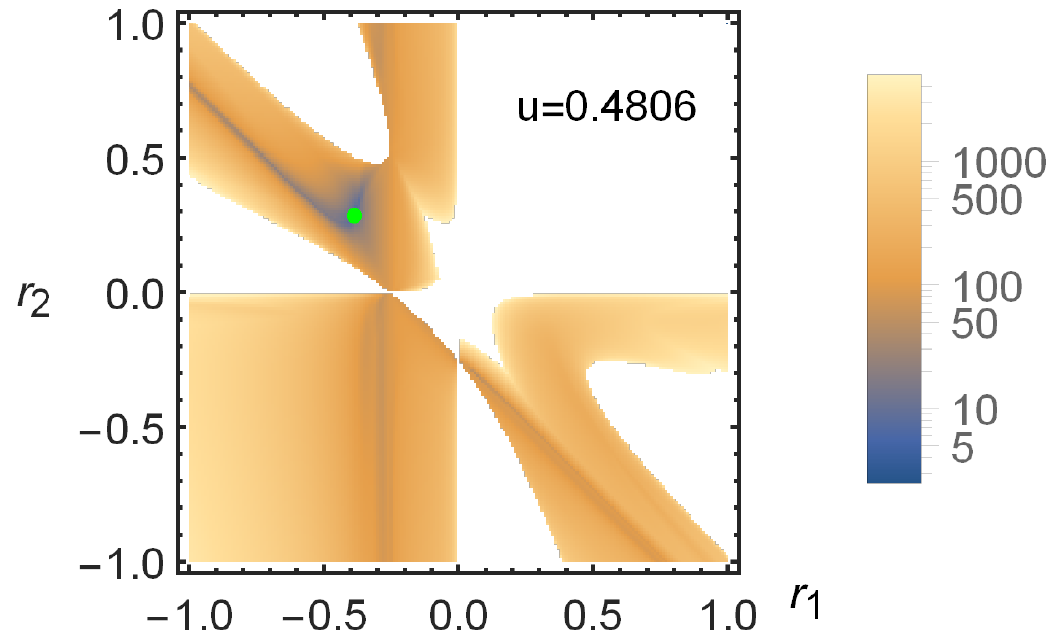}}\\
  \end{tabular}
  \caption{\label{fig:n3Errors}
  Error measures $\veps$ for various decompositions with $n=3$ terms and locations of minimal solutions.
  \textbf{(a)} Second order type-S decompositions with $m=9$ factors. Red: global minimum \eqref{eq:dec-n3p2Best}. At the minimum, the coefficients of the terms $[B,[A,B]]$, $[B,[B,C]]$, and $[C,[A,C]]$ in the expansion \eqref{eq:Uexpansion} vanish.
  \textbf{(b)} Fourth order SE decompositions with $m=17$ factors. Red: global minimum \eqref{eq:dec-n3p4m17SE}. The locations of two 	further minima mentioned in the text are indicated by blue and purple dots.
  \textbf{(c)} Fourth order SE decompositions with $m=21$ factors. Red: global minimum \eqref{eq:dec-n3p4m21SE}. The locations of three further minima mentioned in the text are indicated by green, yellow, and cyan dots.
  \textbf{(d)} Sixth order SL decompositions with $m=37$ factors. Red: global minimum \eqref{eq:dec-n3p6m37SL-1}, blue: second local minimum \eqref{eq:dec-n3p6m37SL-2}.
  \textbf{(e,f)} Fourth order SE decompositions with $m=25$ factors. Red: global minimum \eqref{eq:dec-n3p4m25SE} with $\veps\approx 3.3799$, green: second local minimum with $\veps\approx 6.1855$.
  }
\end{figure}
\myitem
\emph{Leapfrog ($m=5$, type S, $\nu=1$).} -- The only parameter is fixed to $w_1=1$ by the constraint that the first order term in $\log U$ is $tH=t(A+B+C)$. The error is $\veps = \left(\frac{5}{2}\right)^2\frac{13}{24} = \frac{325}{96} \approx 3.3854$ ($A<B$ and $A<C$).

\myitemBest
\emph{Optimized ($m=9$, type S, $\nu=5$).} -- There are three constraints and hence two free parameters; we choose $\{a_1,b_1\}$. The error is minimized for
\begin{equation}\label{eq:dec-n3p2Best}
\begin{split}
	&\ts c_1=\frac{1}{2},\quad b_2=\frac{1}{2}-b_1,\quad a_2=1-2a_1,\quad\\
	&\ts a_1=\frac{1}{6}\approx0.16667,\quad b_1=\frac{1}{6}\left(3-\sqrt{3}\right)\approx 0.21132
\end{split}
\end{equation}
which gives $\veps\approx 1.0496$ ($B<A$ or $C<A$). At this point, the coefficients of the terms $[B,[A,B]]$,
$[B,[B,C]]$, and $[C,[A,C]]$ vanish. See Fig.~\ref{fig:n3Errors}(a).

\myitem
\emph{Optimized ($m=11$, type S-abc, $\nu=6$).} -- There are three constraints and hence three free parameters; we choose $\{a_1,b_1,c_1\}$. The error is minimized for
\begin{equation}
\begin{split}
	&\ts a_2=\frac{1}{2}-a_1,\quad b_2=\frac{1}{2}-b_1,\quad c_2=1-2c_1,\quad a_1=0.098049260850570928723,\\
	&\ts b_1=0.20732225423860549595,\quad c_1=0.35418178737720793097
\end{split}
\end{equation}
which gives $\veps\approx 2.3391$. At this point, the coefficients of the terms $[B,[A,C]]$ and $[C,[A,B]]$ vanish.

\myitem
\emph{Optimized ($m=11$, type S, $\nu=6$).} -- There are three constraints and hence three free parameters; we choose $\{a_1,b_1,b_2\}$. The error is minimized for
\begin{equation}
\begin{split}
	&\ts c_1=\frac{1}{2},\quad a_2=\frac{1}{2}-a_1,\quad b_3=1-2(b_1+b_2),\quad\\
	&\ts a_1=\frac{1}{6}\approx0.16667,\quad b_1=\frac{1}{6} \left(3-\sqrt{3}\right)\approx 0.21132,\quad
	     b_2=\frac{1}{24} \left(4 \sqrt{3}-3\right)\approx 0.16368
\end{split}
\end{equation}
which gives $\veps\approx 1.3054$ ($B<A$ or $C<A$). At this point, the coefficients of the terms $[A,[A,B]]$, $[B,[B,C]]$, and $[C,[A,C]]$ vanish.

\myitemSummary
\emph{Discussion.} -- The best decomposition found here is of type S with $m=9$ as specified in Eq.~\eqref{eq:dec-n3p2Best}. It improves over the common leapfrog decomposition by a factor $\sim 1/3$. According to Table~\ref{tab:n3NoConstraints}, we can reach order $p=4$ with $m=13$ factors.

\subsection{Order \texorpdfstring{$p=4$}{p=4}}
\myitem
\emph{Forest \& Ruth, Yoshida ($m=13$, type SL).} -- For the decomposition $U_{\tY,q=2}$ with $q=2$ in Eq.~\eqref{eq:Yoshida}, the error is $\veps \approx 65.721$ ($A<C<B$).

\myitem
\emph{Optimized ($m=17$, type S-abc, $\nu=9$).} -- There are no free parameters and two real solutions. The first solution reproduces the type-SL decomposition above with $m=13$. The second solution has a very large error $\veps \approx 1243.13$.

\myitem
\emph{Optimized ($m=17$, type SL, $\nu=2$).} -- There are two constraints and hence no free parameters. According to the Gr\"{o}bner basis, there are only two complex solutions.

\myitem
\emph{Optimized ($m=17$, type SE, $\nu=4$).} -- There are three constraints and hence there is one free parameter. For practical reasons, we reparametrize according to
\begin{subequations}\label{eq:dec-n3p4m17SE}
\begin{equation}
	u:=u_1,\quad q_1:=u_1+v_1,\quad r_1:=v_1+u_2,\quad q_2:=u_2+v_2.
\end{equation}
According to the Gr\"{o}bner basis, we can choose $r_1$ as the free parameter. The error is minimized for
\begin{equation}
\begin{split}
	&\ts q_2 =\frac{1}{2}-q_1,\quad u = 0.17981480932806103194,\\
	&\ts r_1=-0.057483169922767706230,\quad q_1=0.73912878293102653974
\end{split}
\end{equation}
which gives $\veps\approx 15.3395$ ($A<B<C$). A nearby analytical solution with almost identical error is
\begin{equation}
\begin{split}
	&\ts q_2 =\frac{1}{2}-q_1,\quad u = \frac{1}{102} \left(57-6 \sqrt{30}-\sqrt{231-36 \sqrt{30}}\right),\\
	&\ts r_1=-\frac{1}{17},\quad
	   q_1= \frac{1}{12} \left(3+\sqrt{231-36 \sqrt{30}}\right).
\end{split}
\end{equation}
There are two further local minima with $\veps\approx 16.3522$ ($B<A<C$) and $\veps\approx 17.420$ ($C<B<A$), respectively. See Fig.~\ref{fig:n3Errors}(b).
\end{subequations}

\myitem
\emph{Suzuki, McLachlan, Kahan \& Li, Omelyan et al.\ ($m=21$, type SL).} -- For Suzuki's decomposition $U_{\tZ,q=2}$ in Eq.~\eqref{eq:Suzuki}, the error is $\veps\approx 35.239 $. For McLachlan's decomposition \eqref{eq:dec-n2p4m11SL-McL} and the decomposition \eqref{eq:dec-n2p4m11SL-Omel} of Omelyan et al., adapted to the $n=3$ case, we find errors $\veps\approx 19.479$ and $\veps\approx 22.827$, respectively. Kahan and Li's decompositions \eqref{eq:dec-n2p4m11SL-KahanLi} have the error $\veps\approx 33.346$ (order $A<B<C$ in all cases). 

\myitem
\emph{Optimized ($m=21$, type SL, $\nu=3$).} -- There are two constraints and hence there is one free parameter; according to the Gr\"{o}bner basis, we can choose $w_2$. The error is minimized for
\begin{subequations}
\begin{equation}\label{eq:dec-n3p4m21SL}
	w_3=1-2(w_1+w_2),\quad w_1=0.25733995540811130577,\quad w_2 =0.6765218865807686
\end{equation}
which gives $\veps\approx 18.968$ ($A<B<C$). A nearby analytical solution with almost identical error is
\begin{equation}\ts
	w_3=1-2(w_1+w_2),\quad w_1=\frac{1}{18} \left(\left(278-6 \sqrt{2145}\right)^{1/3}+\left(278+6\sqrt{2145}\right)^{1/3}-4\right),\quad w_2 =\frac{2}{3}.
\end{equation}
This happens to coincide with the decomposition Eq.~\eqref{eq:dec-n2p4m11SL-1A} for $n=2$ terms.
A second local minimum is located at
\begin{equation}\ts
	w_3=1-2(w_1+w_2),\quad w_1=0.75433412633084310590,\quad w_2 =0.22503541239785228348
\end{equation}
with error $\veps\approx 29.284$ ($A<B<C$). At this point, the coefficient of several terms like  $[B,[B,[B,[A,B]]]]$ vanishes.
\end{subequations}

\myitemBest
\emph{Optimized ($m=21$, type SE, $\nu=5$).} -- There are three constraints and hence two free parameters. For practical reasons, we reparametrize according to
\begin{subequations}\label{eq:dec-n3p4m21SE}
\begin{equation}
	u:=u_1,\quad q_1:=u_1+v_1,\quad r_1:=v_1+u_2,\quad q_2:=u_2+v_2,\quad r_2:=v_2+u_3.
\end{equation}
According to the Gr\"{o}bner basis, we can choose $\{r_1,q_2\}$ as the free parameters. The error is minimized for
\begin{equation}
\begin{split}
	&\ts r_2 = \frac{1}{2}-(u+r_1),\quad u = 0.095968145884398107402,\quad q_1=0.43046123580897338276\\
	&\ts r_1=-0.075403897922216340661,\quad q_2=-0.12443549678124729963
\end{split}
\end{equation}
which gives $\veps\approx 3.92577$ ($C<A<B$). There are three further local minima with $\veps\approx 5.4935$ ($C<A<B$), $\veps\approx 5.6819$ ($B<A<C$), and $\veps\approx 6.9253$ ($C<A<B$). See Fig.~\ref{fig:n3Errors}(c).
\end{subequations}

\myitem
\emph{Optimized ($m=25$, type SL, $\nu=3$).} -- There are two constraints and hence there is one free parameter; according to the Gr\"{o}bner basis, we can choose $w_2$. The error is minimized for
\begin{equation}\label{eq:dec-n3p4m25SL}\ts
	w_3=\frac{1}{2}-(w_1+w_2), \ \
	w_1=\frac{1}{6} \left(2+2^{-1/3}+2^{1/3}\right)\approx 0.6756, \ \
	w_2 =-\frac{1}{6} \left(1+2^{1/3}\right)^2\approx -0.8512
\end{equation}
which gives $\veps\approx 56.179$ ($A<C<B$).
\begin{figure}[t]
  \includegraphics[width=0.94\columnwidth]{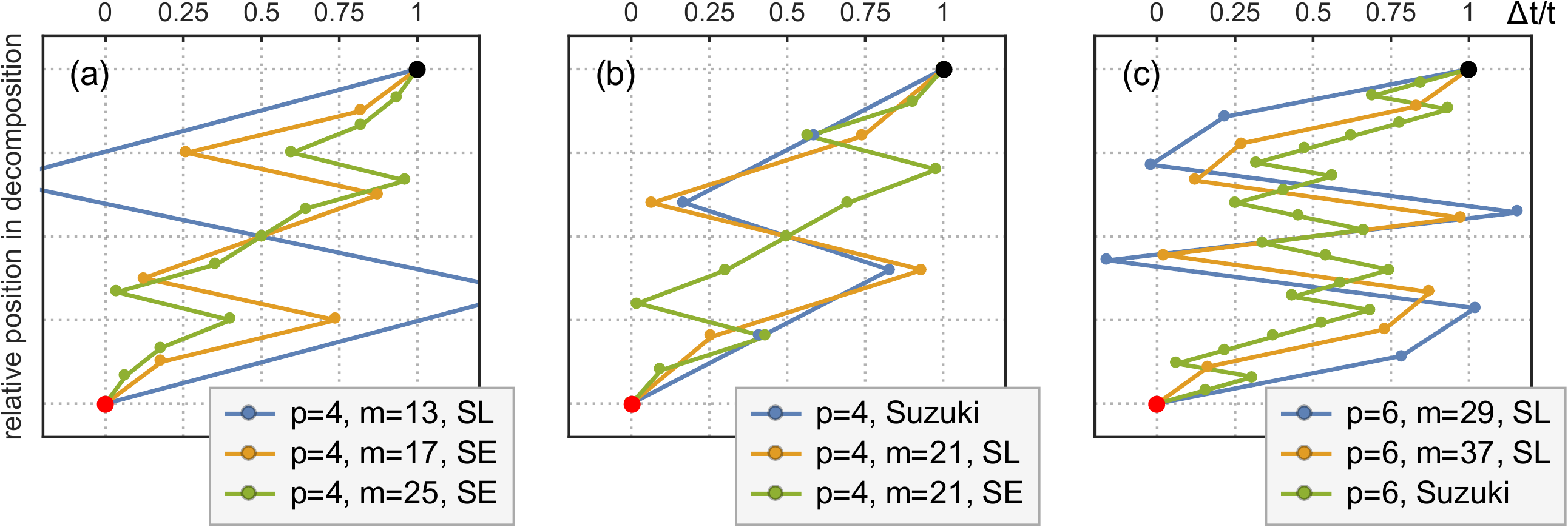}
  \caption{\label{fig:n3Decomps}Lie-Trotter-Suzuki decompositions for $n=3$ terms, i.e., $H=A+B+C$. The shown SE and SL decompositions start at the red dots with $\Delta t/t=0$ and then take Euler steps \eqref{eq:Euler} or leapfrog steps \eqref{eq:leapfrog3} until arriving at the black dot representing $\Delta t/t=1$.
  \textbf{(a)} Yoshida's fourth order decomposition $U_{\tY,q=2}$ in Eq.~\eqref{eq:Yoshida} with the minimal number of $m=13$ factors, and optimized type-SE decompositions \eqref{eq:dec-n3p4m17SE} and \eqref{eq:dec-n3p4m25SE} with $m=17$ and $m=25$, respectively.
  \textbf{(b)} Suzuki's fourth order decomposition $U_{\tZ,q=3}$ in Eq.~\eqref{eq:Suzuki} with $m=21$ factors, and the optimized type-SL and type-SE decompositions \eqref{eq:dec-n3p4m21SL} and \eqref{eq:dec-n3p4m21SE} with $m=21$ factors. The latter is recommended.
  \textbf{(c)} Sixth order type-SL decompositions: decomposition \eqref{eq:dec-n3p6m29SL} with the minimal number of $m=29$ factors, the recommended decomposition \eqref{eq:dec-n3p6m37SL-1} with $m=37$, and Suzuki's decomposition $U_{\tZ,q=3}$ in Eq.~\eqref{eq:Suzuki} with $m=101$.}
\end{figure}

\myitem
\emph{Optimized ($m=25$, type SE, $\nu=6$).} -- There are three constraints and hence there are three free parameters. For practical reasons, we reparametrize according to
\begin{subequations}\label{eq:dec-n3p4m25SE}
\begin{equation}
	u:=u_1,\quad q_1:=u_1+v_1,\quad r_1:=v_1+u_2,\quad q_2:=u_2+v_2,\quad r_2:=v_2+u_3,\quad q_3:=u_3+v_3.
\end{equation}
According to the Gr\"{o}bner basis, we can choose $\{u,r_1,r_2\}$ as the free parameters. The error is minimized for
\begin{equation}
\begin{split}
	&\ts q_3 = \frac{1}{2}-(q_1+q_2),\quad u = 657/10000=0.0657,\\
	&\ts q_1 = \frac{164817921201-1207 \sqrt{186292620253182}}{834300125568}\approx 0.17781,
	\quad r_1=\frac{42}{125}=0.336,\\
	&\ts q_2 = \frac{21225084384-2887 \sqrt{186292620253182}}{128353865472}\approx -0.14163,
	\quad r_2=-\frac{28}{625}=-0.0448
\end{split}
\end{equation}
which gives $\veps\approx 3.3799$ ($B<A<C$). There is another local minimum with $\veps\approx 6.1855$ ($C<A<B$). See Fig.~\ref{fig:n3Errors}(e,f).
\end{subequations}

\myitemSummary
\emph{Discussion.} -- The best decomposition found here is of type SE with $m=25$ as specified in Eq.~\eqref{eq:dec-n3p4m25SE}. It improves over the decomposition $U_{\tY,q=2}$ [Eq.~\eqref{eq:Yoshida}] of Forest, Ruth, and Yoshida \cite{Forest1990-43,Yoshida1990} by a factor $\sim 1/20$ and by a factor of $\sim 1/10$ over the decomposition $U_{\tZ,q=2}$ [Eq.~\eqref{eq:Suzuki}] due to Suzuki \cite{Suzuki1990}. Actually, the error of the decomposition \eqref{eq:dec-n3p4m25SE} with $m=25$ factors does not improve too much over that of the type-SE decomposition with $m=21$ factors [Eq.~\eqref{eq:dec-n3p4m21SE}]. We hence recommend using the latter for fourth order integration.
It is not surprising that the optimized type-SL decomposition \eqref{eq:dec-n3p4m25SL} with $m=25$ factors has a larger error than the SL decomposition \eqref{eq:dec-n3p4m21SL} with $m=21$ factors. The number of free parameters does not increase when going to $m=25$, but the increased number of factors is taken account of in the definition \eqref{eq:n2Error} of $\veps$ and results in a larger error value.
We have checked explicitly that type-S decompositions with $m=13$ factors just reproduce the type-SL decomposition $U_{\tY,q=2}$ due to Forest, Ruth, and Yoshida, and that type-S decompositions with $m=17$ and $m=21$ reproduce the type-SE decompositions \eqref{eq:dec-n3p4m17SE} and \eqref{eq:dec-n3p4m21SE}, respectively. Similarly, all inspected solutions for type-S decompositions with $m=25$ were of type SE.
According to Table~\ref{tab:n3NoConstraints}, we can reach order $p=4$ with $m=29$ factors.

\subsection{Order \texorpdfstring{$p=6$}{p=6}}
\myitem
\emph{Optimized ($m=29$, type SL, $\nu=4$).} -- There are four constraints and hence no free parameters. According to the Gr\"{o}bner basis, there are three real solutions. The best of the three solutions is
\begin{equation}\label{eq:dec-n3p6m29SL}
\begin{split}
	&w_4=1-2(w_1+w_2+w_3),\quad w_1=0.78451361047755726382,\\
	&w_2=0.23557321335935813368,\quad w_3=-1.1776799841788710069
\end{split}
\end{equation}
which has error $\veps \approx 722.85$ ($A<B<C$). It, of course, coincides with the corresponding decomposition \eqref{eq:dec-n2p6m15SL} for $n=2$ terms. The other two solutions have much larger errors $\veps \approx 15940$ and $\veps \approx 16470$, respectively (both for order  $A<B<C$).

\myitem
\emph{Yoshida, Kahan \& Li ($m=37$, type SL).} -- For Yoshida's decomposition $U_{\tY,q=3}$ with $q=3$ in Eq.~\eqref{eq:Yoshida}, the error is $\veps \approx 68024$ ($A<B<C$). Kahan and Li's decomposition \eqref{eq:dec-n2p6m19SL-KahanLi} has the error $\veps\approx 687.06$ ($A<B<C$).

\myitemBest
\emph{Optimized ($m=37$, type SL, $\nu=5$).} -- There are four constraints and hence there is one free parameter; according to the Gr\"{o}bner basis, we can choose $w_4$. The error is minimized for
\begin{subequations}
\begin{equation}\label{eq:dec-n3p6m37SL-1}
\begin{split}
	&w_5=1-2(w_1+w_2+w_3+w_4),\quad w_1=0.16659349375998375835,\\
	&w_2 =0.56336178134626382570,\quad w_3=0.14590936034821488251,\quad w_4=-0.852319424
\end{split}
\end{equation}
which gives $\veps\approx 411.08$ ($A<B<C$). A second local minimum is located at
\begin{equation}\label{eq:dec-n3p6m37SL-2}
\begin{split}
	&w_5=1-2(w_1+w_2+w_3+w_4),\quad w_1=0.30049931385485146980,\\
	&w_2=0.56792684581184873321,\quad w_3=-0.89703459487987352595,\quad w_4=0.024808114
\end{split}
\end{equation}
with error $\veps\approx 571.12$ ($A<B<C$). See Fig.~\ref{fig:n3Errors}(d).
\end{subequations}

\myitem
\emph{Suzuki ($m=101$, type SL).} -- For the decomposition $U_{\tZ,q=3}$ in Eq.~\eqref{eq:Suzuki}, the error is $\veps\approx 51034$ ($A<B<C$).
\begin{figure}[t]
  \includegraphics[height=0.215\columnwidth]{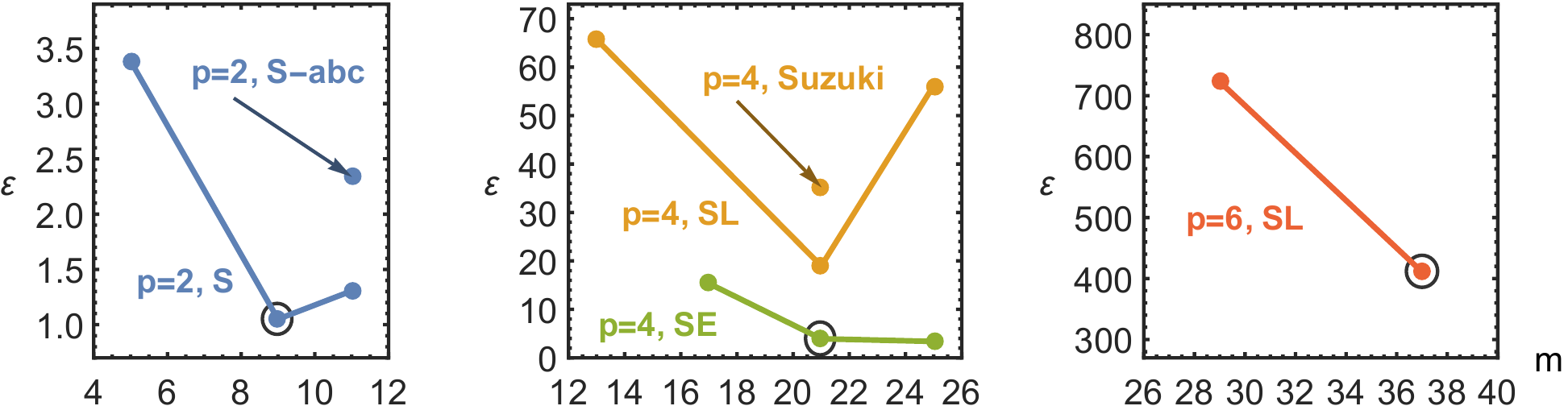}
  \caption{\label{fig:n3_errors}Error values $\veps$ for the optimized $n=3$ decompositions specified in Sec.~\ref{sec:n3opt}, and Suzuki's fourth order decomposition \eqref{eq:Suzuki}. Yoshida's and Suzuki's unoptimized sixth order integrators are not displayed because of their large errors. Recommended decompositions for each order are indicated by circles. As discussed in Sec.~\ref{sec:n2ErrorMeasure}, error values of decompositions with different order $p$ can not be compared directly.}
\end{figure}

\myitemSummary
\emph{Discussion.} --
The best decomposition found here is of type SL with $m=37$ as specified in Eq.~\eqref{eq:dec-n3p6m37SL-1}. It improves over Yoshida's generic decomposition $U_{\tY,q=3}$ [Eq.~\eqref{eq:Yoshida}] by a factor $\sim 1/165$, by a factor of $\sim 1/124$ over Suzuki's decomposition $U_{\tZ,q=3}$ [Eq.~\eqref{eq:Suzuki}], and by a factor of $\sim 4/7$ over the simplest sixth order decomposition \eqref{eq:dec-n3p6m29SL}.
According to Table~\ref{tab:n3NoConstraints}, true type-SE and type-S decompositions require $m=37$ and $m=117$ factors, respectively, to reach order $p=6$, and there exist type-SL decompositions of order $p=8$ for $m\geq 61$ factors.

\section{Discussion}\label{sec:discuss}
We have determined optimized Lie-Trotter-Suzuki decompositions for $n=2$ and $n=3$ terms up to order $t^6$. Using a coarse-graining argument, we have explained why these decompositions are sufficient to simulate any 1d and 2d lattice models with finite-range interactions. Decompositions of different approximation order are constructed by expanding in terms of nested commutators, using Hall bases to remove linear dependencies, and solving systems of polynomial constraints resulting from the comparison with $e^{tH}$. The sizes of Hall bases are also essential to understand the numbers of constraints and free parameters for the different decomposition types. The free parameters are used to minimize the amplitudes of leading error terms. For these optimizations, we employ an error measure that bounds the operator-norm distance and allows for a fair comparison of decompositions with different numbers of factors $m$ in the sense that the time step $t$ should be chosen proportional to $m$ to keep computation costs constant.

For $n=2$ terms, at order $p=2$, we recommend the type-S decomposition \eqref{eq:dec-n2p2Best} with $m=5$ factors, at order $p=4$, the type-S decomposition \eqref{eq:dec-n2p4m11S-1} with $m=11$ factors and, at order $p=6$, the type-SL decomposition \eqref{eq:dec-n2p6m19SL} with $m=19$ factors. We have applied the recommended $p=4$ decomposition, in particular, in many precise tensor network simulations as in Refs.~\cite{Barthel2013-15,Lake2013-111,Cai2013-111,Barthel2016-94,Barthel2017_08unused,Binder2018-98}.

For $n=3$ terms, at order $p=2$, we recommend the type-S decomposition \eqref{eq:dec-n3p2Best} with $m=9$ factors, at order $p=4$, the type-SE decomposition \eqref{eq:dec-n3p4m21SE} with $m=21$ factors and, at order $p=6$, the type-SL decomposition \eqref{eq:dec-n3p6m37SL-1} with $m=37$ factors.

Ref.~\cite{Sornborger1999-60} discusses decompositions for an arbitrary number of terms $n$. For $n=2$ and $n=3$ they have errors similar to those of the decomposition $U_{\tY,q=2}$ in Eq.~\eqref{eq:Yoshida} due to Forest, Ruth, and Yoshida \cite{Forest1990-43,Yoshida1990}. Note that the type-SL and type-SE decompositions presented here are generally applicable for any number of terms $n$, but they are in general not optimal when applied for $n\geq 4$.

Of course there are alternatives to using Lie-Trotter-Suzuki decompositions. Tensor network states and matrix product states, in particular, can also be evolved using Runge-Kutta methods \cite{Cazalilla2001,Feiguin2005}, Krylov subspace methods \cite{Schmitteckert2004-70,Garcia-Ripoll2006-8,Dargel2012-85,Wall2012-14}, or the time-dependent variational principle \cite{Haegeman2011-107,Haegeman2016-96}. Some reviews are given in Refs.\ \cite{Garcia-Ripoll2006-8,Schollwoeck2011-326,Paeckel2019_01}. For the purpose of digital quantum simulation (a.k.a.\ Hamiltonian simulation), algorithms with a gate count that is poly-logarithmic in the desired accuracy have been developed \cite{Berry2014-283,Berry2015-114,Low2017-118,Haah2018_01}, e.g., by introducing ancillary qubits and implementing truncated Taylor expansions.
For classical systems, popular choices are linear multistep methods and Runge-Kutta methods.

We gratefully acknowledge discussions with R.\ Mosseri and J.\ Socolar, and support through US Department of Energy grant DE-SC0019449.

\end{document}